\documentclass[english, aps, prl, reprint, superscriptaddress, longbibliography]{revtex4-1}
\pdfoutput=1
\usepackage[T1]{fontenc}
\usepackage[latin9]{inputenc}
\usepackage{amsmath}
\usepackage{amssymb}
\usepackage{graphicx}
\usepackage{babel}
\usepackage{color}
\usepackage[normalem]{ulem}
\usepackage[caption=false]{subfig}
\usepackage{epsf}
\usepackage{rotating}
\usepackage{epsfig}
\usepackage{dcolumn}
\usepackage{bm}

\def\sumij1{\sum_{t=0}^{t_{\scriptsize{\mbox{max}}}-1}\sum_{i,j=1}^{N}}

\def\cE{{\cal E}}
\def\tmax{T}

\def\density{\rho}
\def\flux{f}
\def\tf{t^{\text{free}}}
\def\densjam{\density^{\text{jam}}}

\newcommand{\etal}[1]{\emph{~et al.}}

\begin{document}
	
	\title{Reducing Urban Traffic Congestion Due To Localized Routing Decisions}
		
	\author{Bo Li}
	\email{b.li10@aston.ac.uk}
	\selectlanguage{english}%
	\affiliation{Non-linearity and Complexity Research Group, Aston University, Birmingham, B4 7ET, United Kingdom}
	
	\author{David Saad}
	\email{d.saad@aston.ac.uk}
	\selectlanguage{english}%
	\affiliation{Non-linearity and Complexity Research Group, Aston University, Birmingham, B4 7ET, United Kingdom}

	\author{Andrey Y. Lokhov}
	\email{lokhov@lanl.gov}
	\selectlanguage{english}%
	\affiliation{Theoretical Division, Los Alamos National Laboratory, Los Alamos, New Mexico 87545, USA}

    \begin{abstract}
     Balancing traffic flow by influencing drivers' route choices to alleviate congestion is becoming increasingly more appealing in urban traffic planning. Here, we introduce a discrete dynamical model comprising users who make their own routing choices on the basis of local information and those who consider routing advice based on localized inducement. We identify the formation of traffic patterns, develop a scalable optimization method for identifying control values used for user guidance, and test the effectiveness of these measures on synthetic and real-world road networks.
    \end{abstract}

	\maketitle

Many of the world's major cities are increasingly gridlocked with a staggering estimated annual cost of \$166B in the United States alone~\cite{tamu2019}. Relentless urban population growth has created exorbitant traffic demands, which leads to recurring large-scale traffic jams~\cite{Gorzelany2013, DfT2018, Akorede2019}. Since it is expensive to satisfy the demand exclusively through further investment in infrastructure, there is a growing interest in optimizing transportation systems within the existing infrastructure~\cite{Peeta2001, Hamilton2013, Colak2016}. Modern information technologies can potentially offer effective solutions through ride sharing using smart phones~\cite{AlonsoMora2017}, congestion-aware routing schemes~\cite{Lim2011}, and the use of autonomous vehicles~\cite{Thrun2006, Fagnant2015}. The deployment of smart devices already impacts transportation networks, leading to a paradigm shift in traffic planning and management. However, not all these changes are for the better. Most navigation apps have been designed typically to minimize an individual driver's travel time irrespective of street capacity along the route, safety, or the route choices of other drivers; in many times this results in traffic chaos~\cite{Macfarlane2019}. Moreover, recent simulation results demonstrate the potential of having a mixed environment, of drivers who make their own route choices en route and those who follow routing advice that is centrally optimized, in reducing congestion~\cite{pmlr-v78-wu17a}; this scenario is inherently accommodated within the framework presented here. It is therefore important to understand the potential and limitations of these technologies and develop scalable algorithmic tools that would enable their use in real-time settings.
 
Detailed microscopic modeling of multiagent systems characterizing the paths of individual users, such as cellular automata-based simulations~\cite{Nagel1992}, model basic traffic systems but usually require considerable computational power; it is also generally difficult to gain insight due to the overwhelming level of details. On the other hand, models based on traffic flow, that coarse-grain the behaviors of individual users but maintain correlations at the network level, are simplistic but amenable to analysis. Link-based methods have been developed along this line, mainly for static assignments, selfish routing or centralized optimization~\cite{Bar-Gera1999, Roughgarden2005}. Such methods have also been extended to the more difficult dynamic traffic assignment problem~\cite{Szeto2012}; for instance, the Wardrop's static equilibrium principle was extended to dynamic scenarios~\cite{Friesz1989}. In reality, drivers do not have full information of the traffic flow and unbounded computational capacity to determine the rational route-choices~\cite{Ben-Elia2010, Ben-Elia2013}. Instead, they typically adjust their route choice, especially in urban settings, en route according to the traffic conditions in downstream junctions, which has been investigated in some dynamic traffic assignment problems~\cite{BenAkiva1991, Kuwahara1997, Pel2009}. 
    
In this Rapid Communication, we take into account such behavioral aspects and propose a dynamical model which includes both impulsive users who make their own decisions en route, and advice-susceptible users who follow the suggestions given by smart devices. Advice-susceptible users are incentivized to follow centrally optimized routing suggestions that benefit traffic globally. Such a strategy may be adopted in the future to alleviate traffic congestion~\cite{pmlr-v78-wu17a}. In fact, electronic road pricing already operates successfully in Singapore~\cite{ERP_Singapore}, and has been recently launched in Israel to motivate drivers into driving in nonrush hours and carpooling~\cite{Barak2019}. Our computational model offers complementary insights in support of such strategies. We focus on scenarios where commuters travel towards the city center at peak hours, during which they typically experience severe traffic congestion. A realistic model of this type is naturally nonlinear, and hard to optimize; one of the contributions of this Rapid Communication consists in developing a scalable and computationally efficient optimization method, that supports real-time applications. We analyze the characteristics of emerging traffic patterns, develop an algorithm to determine the optimal incentive and investigate their impact on traffic congestion.

We model the urban road system as a network, where intersections are mapped to nodes and roads between them to edges (or links). We consider a scenario where drivers travel towards a universal destination $D$, which is relevant in the morning rush hour when a large number of people commute to the city center. The network is depicted as an undirected graph $G(V, E)$ of $N$ nodes, where each node $i \in V$ is connected to $k_i$ neighbors denoted by $\partial i$, and each edge $(i,j) \in E$ represents two lanes $i \to j$ and $j \to i$, accommodating non-interacting traffic from $i$ to $j$ and $j$ to $i$, respectively. We denote the set of all lanes as $\cE$.
	
Assume that drivers can be classified into two groups according to whether they make their own route choices or follow the advice from navigation devices. In the former, a user makes routing decisions dynamically, based on her estimated time to destination $D$. Upon arriving at intersection $i$ at time $t$, the user faces a choice between $k_i$ possible roads $\{ i \to j \}^{k_i}_{j=1}$. The user first estimates (i) the time it takes to travel through edge $i \to j$ as $g(\density^{t}_{ij})$ where $\density^{t}_{ij}$ is the number of users occupying edge $i \to j$ (i.e., traffic volume) at that time and $g(\density^{t}_{ij})$ is determined by the Greenshields model~\cite{Greenshields1936} [see also the Supplemental Material (SM)~\cite{Li2020sup}], and (ii) the remaining time $d_j$ needed to travel to $D$ from node $j$, which can be taken as the shortest free traveling time or be based on past experience of the congestion level. Afterwards, their route choices are made according to the probability
	\begin{equation}
	    p^{g,t}_{ij} (\density^{t}) = \frac{ e^{-\beta [g(\density^{t}_{ij}) + d_j]} }{ \sum_{k \in \partial i} e^{-\beta [g(\density^{t}_{ik}) + d_k]} } \label{eq:def_pg},
	\end{equation}
where $\beta$ is a parameter determining the randomness of the decision making process. As shown in the SM~\cite{Li2020sup}, the dependence on $\beta \geq 1$ is relatively weak, and hence we choose $\beta = 1$ in what follows. Note that we do not limit users from turning back. The awareness of congestion can be extended to road segments that are more distant, at the cost of higher computational complexity. Here, we focus on the one-step congestion-aware model. 
	
In the latter group, users follow the navigation advice aimed at improving traffic efficiency. Their route choices at junction $i$ at time $t$ are determined by the localized probability 
	\begin{equation}
	    p^{w,t}_{ij} (w^{t}) = e^{-w^{t}_{ij}} / \sum_{k \in \partial i} e^{-w^{t}_{ik}}, \label{eq:pw} 
	\end{equation}
where the weight variables $\{ w^{t}_{ij} \}$ are optimized centrally.
With the assumption that the fraction of users $n$ who are susceptible to routing advice are distributed evenly in the network, on average the vehicle flow arriving at node $i$ at time $t$ will be diverted to the adjacent edges $\{ i \to j \}^{k_i}_{j=1}$ according to the distribution
	\begin{equation}
	    p^{t}_{ij} (\density^{t}, w^{t}) = (1-n) p^{g,t}_{ij} (\density^{t}) + n p^{w,t}_{ij} (w^{t}). \label{eq:pt}
	\end{equation}
A similar decision rule has been used to investigate the effect of altruistic users in the \emph{static routing game setting}~\cite{Chen2008, Colak2016}, which differs from the current \emph{dynamical} model.
	
At each time-step $t'$ a decision is made to enter edge $i \to j$, the user then spends some time $\tau_{ij}$ traveling on this edge with distribution $P(\tau_{ij})$, arriving at the end point at time $t=t'+\tau_{ij}$. The distribution of time spent can take several forms, including the typically used discrete Poisson distribution adopted here~\cite{Li2020sup}. The arrival probability depends on the traffic volume $\density^{t'}_{ij}$ at the time of entrance $t'$, i.e. $P(t-t'|\density^{t'}_{ij})$, which is a realistic and an important factor in traffic modeling.
To express the dynamics we introduce the time-dependent flux $\flux^{t}_{ij}$ arriving at the end point $j$ of the edge $i \to j$ at time $t$. Assuming users enter the road system at time $t=0$ with initial volume $\density^{0}$, the dynamics of the traffic volume and flux on edge $i \to j$ ($i \neq D$) are governed by the discrete forward dynamics
	\begin{eqnarray}
	    \density^{t}_{ij} &=& p^{t-1}_{ij}\sum_{ k\in\partial i, k \neq D} \flux^{t-1}_{ki} + (\density^{t-1}_{ij} - \flux^{t-1}_{ij}), \label{eq:forward1} \\
	    \flux^{t}_{ij} &=& \sum^{t}_{t'=1}  \big[ \density^{t'}_{ij} \!-\! (\density^{t'-1}_{ij} \!-\! \flux^{t'-1}_{ij}) \big] P(t-t'|\density^{t'}_{ij}) \nonumber \\
	    && + \density^{0}_{ij} P(t|\density^{0}_{ij}).  \label{eq:forward2}
    \end{eqnarray}
Equation~(\ref{eq:forward1}) describes the traffic volume at edge $i \to j$ at time $t$; it is composed of the newly joined users who selected this junction at node $i$ at time $t-1$, and users who were already traveling through this edge but have not yet reached the end point $j$. Equation~(\ref{eq:forward2}) states that the vehicles flux at the edge $i \to j$ end point at time $t$ comprises the fraction of traffic volume $\density^{t'}_{ij}- (\density^{t'-1}_{ij} - \flux^{t'-1}_{ij})$ entering the road segment at $t'$, who have completed the trip on this road segment within a duration $t-t'$ as dictated by the probability $P(t-t'|\density^{t'}_{ij})$, which is defined such that the mean traveling time follows the Greenshields model~\cite{Greenshields1936, Li2020sup}. The resulting model bears similarity to certain link-based models of dynamic traffic assignment~\cite{Szeto2012}. We assume that no vehicles leave the destination node, i.e., the destination $D$ is an absorbing boundary which satisfies $\density^{t}_{Dj} = \flux^{t}_{Dj} = 0, \forall j \in \partial D$.

The model is simulated for a fixed time window $\tmax$. To evaluate the efficiency of the system, we measure the average time to destination $D$ ahead of $\tmax$,
    \begin{equation}
	    {\cal O} = \frac{1}{\sum_{e \in \cE} \density^{0}_{e} }\sum_{t=1}^{\tmax} (\tmax-t) \sum_{j \in \partial D} \flux^{t}_{jD}, \label{eq:objective}
    \end{equation}
and use it as the main performance measure. Other measures can be easily accommodated within the same framework but will not be considered here.

\begin{figure}
        \includegraphics[scale=0.3]{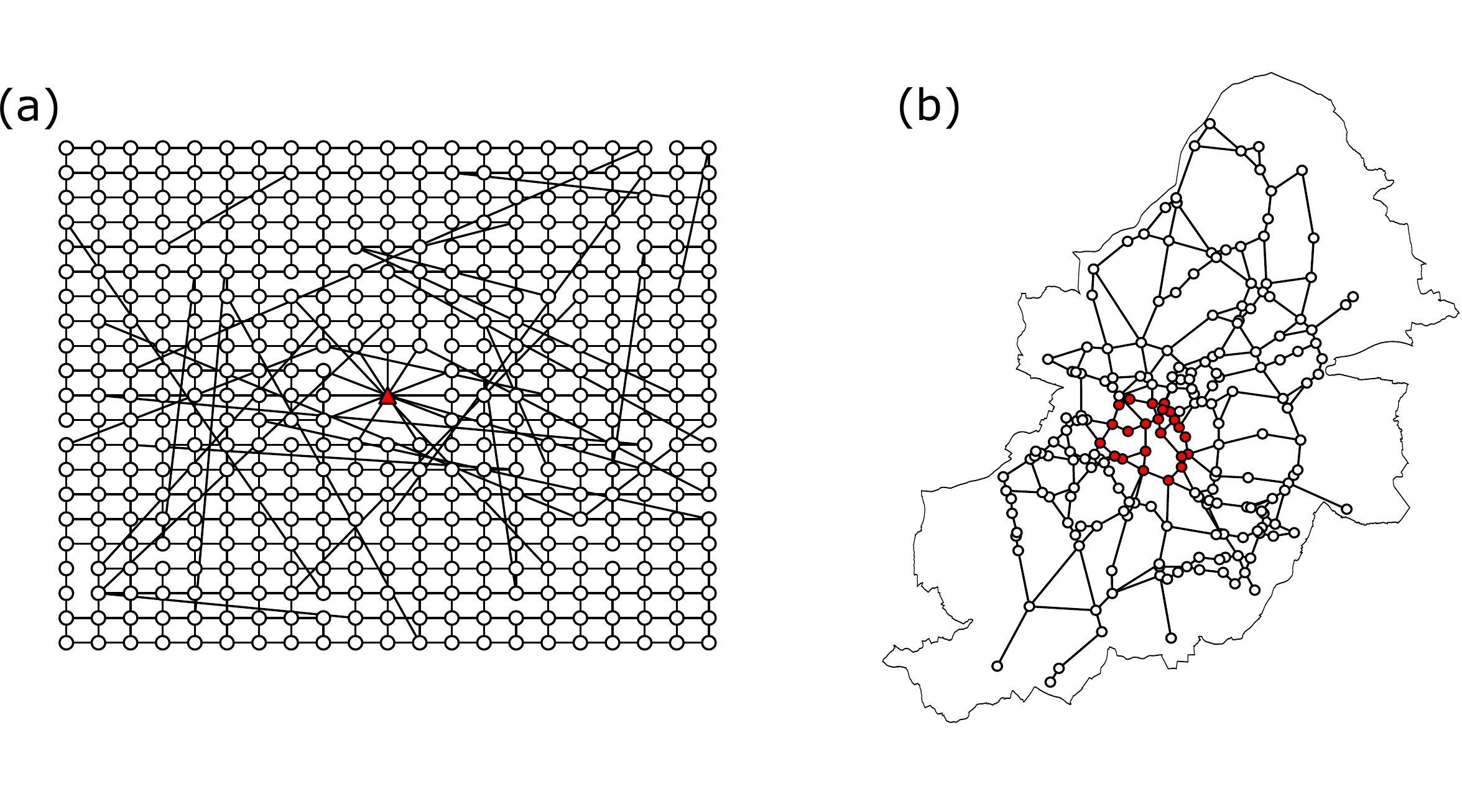}
        \caption{(a) A small-world network generated by rewiring a $21 \times 21$ square lattice with shortcut links with rewiring probability $p_{r}=0.05$. (b) The Birmingham road network composed of major roads in the city of Birmingham, U.K. We define the city center as the the region enclosed by ring road A4540. In both cases, the red nodes constitute the city center and determine the destination in the model.}
        \label{fig:road_networks}
    \end{figure}

We perform numerical experiments on both generated and realistic road networks. The former are constructed by randomly rewiring a planar square lattice with shortcut edges, which is motivated by the recent observation that high-speed urban roads constitute effective long-range connections and render the system to exhibit small-world characteristics~\cite{Zeng2019}. The realistic road network used is extracted from the OpenStreetMap data set~\cite{OpenStreetMap}, and converted to a network format by using the GIS F2E software~\cite{Karduni2016}. Two examples of the networks considered are shown in Fig.~\ref{fig:road_networks}. Details of the network generation are described in the SM~\cite{Li2020sup}.
    
\textit{Model characterization without control.}
The initial traffic volume is assigned independently and identically at random as $\density^{0}_{i}$ users departing from each node $i$, which can be proportional to the population at that node; users rest on the node's neighboring edges $\{i \to j | j \in \partial i\}$ with equal probability $\density^{0}_{ij}=\density^{0}_{i}/k_{i}$, constituting the initial traffic volume $\{ \density^{0}_{ij} \}$. After entering the system at time $t=0$, all users drive towards the center node $D$ according to the instantaneous decision making rule of Eq.~(\ref{eq:def_pg}), i.e., $n=0$ in Eq.~(\ref{eq:pt}). Clearly, the same framework can accommodate users entering the network at any time. It leads to macroscopic dynamical traffic patterns governed by Eqs.~(\ref{eq:forward1}) and~(\ref{eq:forward2}). We define the traffic load level as
    \begin{equation}
        L = \frac{ \sum_{e \in \cE} \density^{0}_{e} }{ \sum_{e \in \cE} \densjam_{e} }.
    \end{equation}
The load level $L$ is similar to the demand-to-supply ratio introduced in Ref.~\cite{Colak2016}, which is suggested to be a good predictor of the congestion level. 

\begin{figure}
        \includegraphics[scale=0.88]{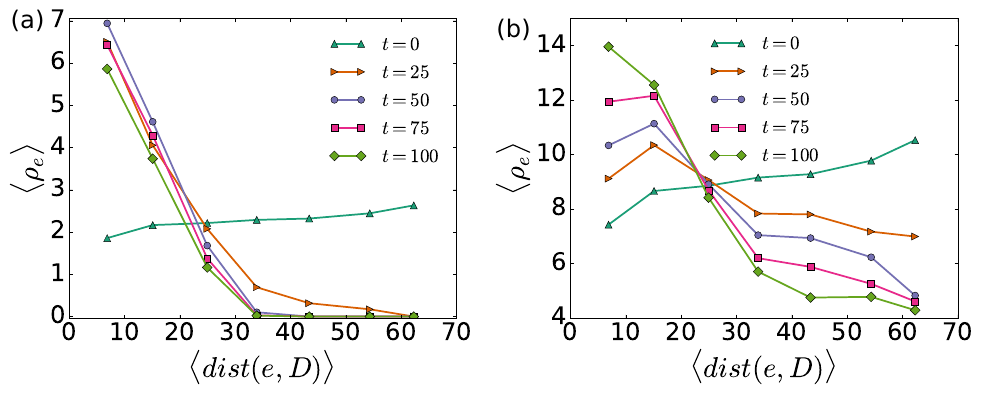}
      \caption{Average volume $\langle \density^{t}_{e} \rangle$ vs average distance to destination $\text{dist}(e, D)$ in a $21 \!\times\! 21$ small-world network. Maximal time is $T\!=\!100$ and users are unaided $n\!=\!0$. The road sections are first binned into groups according to $\text{dist}(e, D)$ in the interval of 10, after which $\density^{t}_{e}$ and $\text{dist}(e, D)$ are averaged within each group. (a) Load level $L\!=\!0.1$. (b) Load level $L\!=\!0.4$.}
        \label{fig:spacetime_evolution}
    \end{figure}

We first study emerging traffic patterns in the absence of routing advice, $n=0$. The movement of traffic mass can be visualized by contrasting the traffic volume $\density_{e}$ to the distance to destination of each lane. To this end, we define the distance $\text{dist}(e, D)$ of lane $e\!=\!(i \to j)$ to destination $D$ as the shortest free traveling time from the midpoint of the lane to destination $\text{dist}(e, D) \!=\! d_{j} \!+\! \tf_{ij}/2$. Fig.~\ref{fig:spacetime_evolution} demonstrates how the average traffic volumes $\langle \density^{t}_{e} \rangle$ at specific distances change over time under two different load levels. At the low load regime $L\!=\!0.1$, shown in Fig.~\ref{fig:spacetime_evolution}(a), the vehicles are able to move fairly quickly towards the destination $D$ from the initial positions at $t\!=\!0$ to $t\!=\!25$. The roads near the city center become congested, leading to a slow clearance of traffic from $t\!=\!50$ to $t\!=\!100$, which indicates that the limited connectivity of the city center is a bottleneck of the traffic system. At high loads $L\!=\!0.4$, shown in Fig.~\ref{fig:spacetime_evolution}(b), the traffic volumes at large distance to destination decrease, while those at short distances increase over time, but at a much slower rate compared to the case of a small load $L\!=\!0.1$. It indicates that the excessive demand creates congestion in the transportation network and leads to an increase in travel time. More details of the system efficiency as a function of load are depicted in the SM~\cite{Li2020sup}.

\begin{figure}
        \centering
        \includegraphics[scale=0.88]{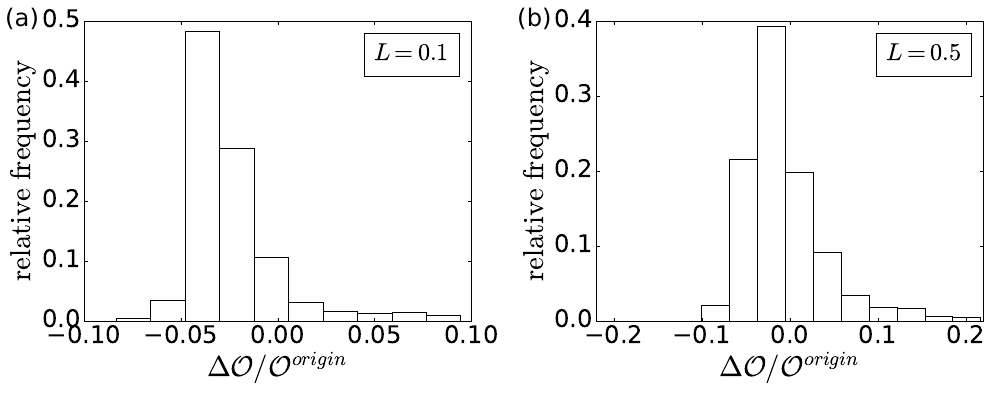}
        \caption{Relative frequency of fractional change of the objective function ${\cal O}$ after adding a link to the existing network, defined as the performance change $\Delta {\cal O} = {\cal O}^{\text{add-link}} - {\cal O}^{\text{origin}}$ divided by the objective function before adding a link ${\cal O}^{\text{origin}}$. The results are aggregated from ten networks generated from the small-world network model of size $21 \times 21$ with rewiring probability $p_r = 0.05$. The parameters are $T=100, n=0$. One end of the new link is randomly chosen from the top five sites with the highest population, while the other end is randomly chosen from the nearest neighbors of the destination node. (a) Load level $L=0.1$. (b) Load level $L=0.5$.}
        \label{fig:add_link_hist}
    \end{figure}

The simplest measure to reduce congestion is to improve the infrastructure, e.g., by building new roads or by increasing the capacity of existing ones. In particular, increasing the number of possible routes to the city center/destination node can significantly enhance the traffic clearance rate, yet it is rarely possible to do so due to the limited land availability. To examine the effect of network extension, we perform experiments by adding links from sites with the largest populations to nearest neighbors of the destination node. From the relative frequency of the fractional change of objective function shown in Fig.~\ref{fig:add_link_hist}, it is surprising to observe that the majority of link additions lead to a \emph{decrease in system performance}. It suggests that newly introduced shortcuts, being attractive to users, create congestion in the shortcut edges and nearby areas. The phenomenon is reminiscent of Braess's paradox in the static routing game~\cite{Braess1968} and other complex systems~\cite{Cohen1991, Witthaut2012, Donovan2018}, where adding resources can possibly lead to a degradation of system performance. In our model, drivers have limited knowledge and are unaware of the long-distance traffic condition, so that the myopic decisions make the system more prone to congestion. If users are aware of more global information as in routing game scenarios, it is possible that they may adapt, in a repeated game scenario, to avoid the already congested shortcuts, such that the probability of performance decrease becomes smaller. 
    
Nevertheless, there is a small likelihood that adding a new link would lead to a significant improvement of the objective function ${\cal O}$, which can be up to $10\%$ for $L=0.1$ and $20\%$ for $L=0.5$. Such an improvement is more commonly observed in higher loads, but in the majority of cases the improvement is marginal. In either case, it is crucial to select the correct shortcut to invest in, which becomes a difficult task when the demand profile is fluctuating.

\textit{Model characterization with control.}
It becomes increasingly more appealing and cost effective to influence the route choice of drivers in order to reduce congestion. We examine the particular type of instantaneous advice in the form of Eq.~(\ref{eq:pw}), which is adapted such that the objective function ${\cal O}$ of Eq.~(\ref{eq:objective}) is maximized. The resulting highly nonlinear optimization problem is nonconvex and suffers from multiple local maxima.

\begin{figure}
        \includegraphics[scale=0.88]{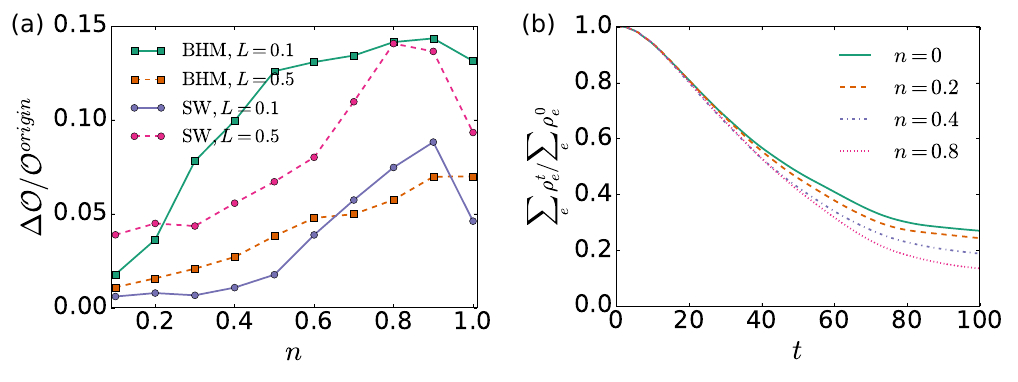}
       \caption{(a) Fractional change in objective function ${\cal O}$ (defined as the performance change $\Delta {\cal O} = {\cal O}^{\text{optimised}} - {\cal O}^{\text{origin}}$ divided by the objective function without advice-susceptible users ${\cal O}^{\text{origin}}$) as a function of the fraction of advice-susceptible users $n$. The Birmingham road network (BHM) and a small world network (SW) of size $21\times 21$ are considered. (b) Time evolution of the fraction of traffic volume $\sum_e \density^t_e / \sum_e \density^0_e$ remaining on the Birmingham road network; at time $t=0$, the system load is $L=0.1$. }
        \label{fig:optimisation}
    \end{figure}

To solve this difficult optimization problem, we adopt an optimal control framework \cite{Chernousko1982, Lokhov2017}, whereby the dynamics, Eqs.~(\ref{eq:forward1}) and (\ref{eq:forward2}), is enforced as constraints in the Lagrangian formulation. The optimality conditions lead to a set of coupled nonlinear equations solved by forward-backward iterations. To suppress divergent behavior due to radical changes of the control parameters~\cite{Chernousko1982, QianxiaoLi2017}, we employ a gradient ascent in the updates of the control parameters~\cite{Li2020sup}. Our method achieves similar objective function values to state-of-the-art constrained optimization approaches, while offering \emph{significant stability and scalability advantages}. This point is illustrated in the SM through a benchmarking comparison to the state-of-the nonlinear programming solver IPOPT~\cite{Li2020sup}.
The results shown in Fig.~\ref{fig:optimisation} demonstrate that the optimization algorithm successfully improves the system performance, as indicated by the fractional increase in the objective function ${\cal O}$ compared to the value ${\cal O}^{\text{origin}}$ without advice-susceptible users. The maximal average improvements are remarkably significant and range from $7\%$ to $14\%$, depending on the network structure and load level. Naively, one would expect for the objective function to monotonically increase with $n$. However, it seems not to be the case in the experiment shown in Fig.~\ref{fig:optimisation}(a) and a slight decrease in performance is shown close to $n=1$. For $n<1$, the mixture probability $p^{t}_{ij} = (1-n) p^{g,t}_{ij} + n p^{w,t}_{ij} (w^{t})$ includes information on the unguided users and effective distance to destination through $p^{g,t}_{ij}$, which facilitates the search for an optimal solution. This information is gradually lost at high $n$ values, resulting in a less pronounced increased performance, compared to the maximally achieved level of gain.
    
In Fig.~\ref{fig:optimisation}(b), we demonstrate the evolution of the fraction of traffic  $\sum_e \density^t_e / \sum_e \density^0_e$ remaining on the Birmingham road network at a given time as a function of the guided-users fraction $n$. One can observe a faster rate of traffic decrease when $n$ increases from 0 to 0.8, suggesting more users can reach the destination within the same time period with the increase in the number of advice-susceptible users.
    
Modeling the dynamics of a transportation network that accommodates different driver behaviors facilitates a greater understanding of the emerging traffic patterns in a regime that is of great interest and relevance~\cite{pmlr-v78-wu17a}, while the suggested optimization scheme provides a scalable and efficient way to implement it, providing better performance than state-of-the-art continuous optimization solvers and offering significant advantages in the online setting, where sudden changes in traffic conditions can be adjusted by a few update steps to obtain a quality approximate solution.  
	We demonstrate how extending the network may result in \emph{increased} congestion and a degradation in system performance, highlighting the importance of a careful selection of the most beneficial roads to add, which will be the subject of future research. Balancing the traffic flow by influencing user  route choices offers a less costly and more flexible solution to the congestion problem.  Our experiments on macroscopic traffic-flow optimization by giving instantaneous and localized routing advice demonstrates its potential for improvements in system performance. The framework also allows for the study of balancing demand by scheduling departure times, which could be integrated into our optimization framework; this is one of the future directions for a follow-up study. These extensions can be tested at a low computational cost using our model and optimization method without the need for expensive large-scale agent-based simulations. Other possible generalizations include the introduction of a spill-back mechanism, the integration of more nonlocal traffic condition information and cases of multiple destinations.

\begin{acknowledgments}
The map data are copyrighted by OpenStreetMap contributors and is available from https://www.openstreetmap.org. B.L. and D.S. acknowledge support from the Leverhulme Trust (RPG-2018-092), European Union's Horizon 2020 research and innovation programme under the Marie Sk{\l}odowska-Curie Grant Agreement No. 835913. D.S. acknowledges support from the EPSRC programme grant TRANSNET (EP/R035342/1). A.Y.L. acknowledges support from the Laboratory Directed Research and Development program of Los Alamos National Laboratory under Projects No. 20190059DR and No. 20200121ER.
\end{acknowledgments}



\bibliography{reference}

\begin{thebibliography}{40}%
\makeatletter
\providecommand \@ifxundefined [1]{%
 \@ifx{#1\undefined}
}%
\providecommand \@ifnum [1]{%
 \ifnum #1\expandafter \@firstoftwo
 \else \expandafter \@secondoftwo
 \fi
}%
\providecommand \@ifx [1]{%
 \ifx #1\expandafter \@firstoftwo
 \else \expandafter \@secondoftwo
 \fi
}%
\providecommand \natexlab [1]{#1}%
\providecommand \enquote  [1]{``#1''}%
\providecommand \bibnamefont  [1]{#1}%
\providecommand \bibfnamefont [1]{#1}%
\providecommand \citenamefont [1]{#1}%
\providecommand \href@noop [0]{\@secondoftwo}%
\providecommand \href [0]{\begingroup \@sanitize@url \@href}%
\providecommand \@href[1]{\@@startlink{#1}\@@href}%
\providecommand \@@href[1]{\endgroup#1\@@endlink}%
\providecommand \@sanitize@url [0]{\catcode `\\12\catcode `\$12\catcode
  `\&12\catcode `\#12\catcode `\^12\catcode `\_12\catcode `\%12\relax}%
\providecommand \@@startlink[1]{}%
\providecommand \@@endlink[0]{}%
\providecommand \url  [0]{\begingroup\@sanitize@url \@url }%
\providecommand \@url [1]{\endgroup\@href {#1}{\urlprefix }}%
\providecommand \urlprefix  [0]{URL }%
\providecommand \Eprint [0]{\href }%
\providecommand \doibase [0]{http://dx.doi.org/}%
\providecommand \selectlanguage [0]{\@gobble}%
\providecommand \bibinfo  [0]{\@secondoftwo}%
\providecommand \bibfield  [0]{\@secondoftwo}%
\providecommand \translation [1]{[#1]}%
\providecommand \BibitemOpen [0]{}%
\providecommand \bibitemStop [0]{}%
\providecommand \bibitemNoStop [0]{.\EOS\space}%
\providecommand \EOS [0]{\spacefactor3000\relax}%
\providecommand \BibitemShut  [1]{\csname bibitem#1\endcsname}%
\let\auto@bib@innerbib\@empty
\bibitem [{\citenamefont {Schrank}\ \emph {et~al.}(2019)\citenamefont
  {Schrank}, \citenamefont {Eisele},\ and\ \citenamefont {Lomax}}]{tamu2019}%
  \BibitemOpen
  \bibfield  {author} {\bibinfo {author} {\bibfnamefont {David}\ \bibnamefont
  {Schrank}}, \bibinfo {author} {\bibfnamefont {Bill}\ \bibnamefont {Eisele}},
  \ and\ \bibinfo {author} {\bibfnamefont {Tim}\ \bibnamefont {Lomax}},\
  }\href@noop {} {\enquote {\bibinfo {title} {2019 urban mobility report},}\
  }\bibinfo {howpublished} {\url{https://mobility.tamu.edu/umr/report/}}
  (\bibinfo {year} {2019})\BibitemShut {NoStop}%
\bibitem [{\citenamefont {Gorzelany}(2013)}]{Gorzelany2013}%
  \BibitemOpen
  \bibfield  {author} {\bibinfo {author} {\bibfnamefont {Jim}\ \bibnamefont
  {Gorzelany}},\ }\href@noop {} {\enquote {\bibinfo {title} {The world's most
  traffic congested cities},}\ }\bibinfo {howpublished} {Forbes,
  \url{https://www.forbes.com/sites/jimgorzelany/2013/04/25/the-worlds-most-traffic-congested-cities/}}
  (\bibinfo {year} {2013})\BibitemShut {NoStop}%
\bibitem [{\citenamefont {for Transport}(2018)}]{DfT2018}%
  \BibitemOpen
  \bibfield  {author} {\bibinfo {author} {\bibfnamefont {Department}\
  \bibnamefont {for Transport}},\ }\href@noop {} {\enquote {\bibinfo {title}
  {Transport statistics great britain: 2018},}\ }\bibinfo {howpublished}
  {\url{https://assets.publishing.service.gov.uk/government/uploads/system/uploads/attachment_data/file/795336/tsgb-2018.pdf}}
  (\bibinfo {year} {2018})\BibitemShut {NoStop}%
\bibitem [{\citenamefont {Akorede}(2019)}]{Akorede2019}%
  \BibitemOpen
  \bibfield  {author} {\bibinfo {author} {\bibfnamefont {Shakir}\ \bibnamefont
  {Akorede}},\ }\href@noop {} {\enquote {\bibinfo {title} {Employees in lagos
  are stressed, burned out and exhausted because of 'hellish traffic'},}\
  }\bibinfo {howpublished} {CNN,
  \url{https://edition.cnn.com/travel/article/traffic-stress-lagos-nigeria/index.html}}
  (\bibinfo {year} {2019})\BibitemShut {NoStop}%
\bibitem [{\citenamefont {Peeta}\ and\ \citenamefont
  {Ziliaskopoulos}(2001)}]{Peeta2001}%
  \BibitemOpen
  \bibfield  {author} {\bibinfo {author} {\bibfnamefont {Srinivas}\
  \bibnamefont {Peeta}}\ and\ \bibinfo {author} {\bibfnamefont {Athanasios~K.}\
  \bibnamefont {Ziliaskopoulos}},\ }\bibfield  {title} {\enquote {\bibinfo
  {title} {Foundations of dynamic traffic assignment: The past, the present and
  the future},}\ }\href {\doibase 10.1023/A:1012827724856} {\bibfield
  {journal} {\bibinfo  {journal} {Networks and Spatial Economics}\ }\textbf
  {\bibinfo {volume} {1}},\ \bibinfo {pages} {233--265} (\bibinfo {year}
  {2001})}\BibitemShut {NoStop}%
\bibitem [{\citenamefont {Hamilton}\ \emph {et~al.}(2013)\citenamefont
  {Hamilton}, \citenamefont {Waterson}, \citenamefont {Cherrett}, \citenamefont
  {Robinson},\ and\ \citenamefont {Snell}}]{Hamilton2013}%
  \BibitemOpen
  \bibfield  {author} {\bibinfo {author} {\bibfnamefont {Andrew}\ \bibnamefont
  {Hamilton}}, \bibinfo {author} {\bibfnamefont {Ben}\ \bibnamefont
  {Waterson}}, \bibinfo {author} {\bibfnamefont {Tom}\ \bibnamefont
  {Cherrett}}, \bibinfo {author} {\bibfnamefont {Andrew}\ \bibnamefont
  {Robinson}}, \ and\ \bibinfo {author} {\bibfnamefont {Ian}\ \bibnamefont
  {Snell}},\ }\bibfield  {title} {\enquote {\bibinfo {title} {The evolution of
  urban traffic control: changing policy and technology},}\ }\href {\doibase
  10.1080/03081060.2012.745318} {\bibfield  {journal} {\bibinfo  {journal}
  {Transportation Planning and Technology}\ }\textbf {\bibinfo {volume} {36}},\
  \bibinfo {pages} {24--43} (\bibinfo {year} {2013})}\BibitemShut {NoStop}%
\bibitem [{\citenamefont {{\c{C}}olak}\ \emph {et~al.}(2016)\citenamefont
  {{\c{C}}olak}, \citenamefont {Lima},\ and\ \citenamefont
  {Gonz{\'a}lez}}]{Colak2016}%
  \BibitemOpen
  \bibfield  {author} {\bibinfo {author} {\bibfnamefont {Serdar}\ \bibnamefont
  {{\c{C}}olak}}, \bibinfo {author} {\bibfnamefont {Antonio}\ \bibnamefont
  {Lima}}, \ and\ \bibinfo {author} {\bibfnamefont {Marta~C.}\ \bibnamefont
  {Gonz{\'a}lez}},\ }\bibfield  {title} {\enquote {\bibinfo {title}
  {Understanding congested travel in urban areas},}\ }\href
  {https://doi.org/10.1038/ncomms10793} {\bibfield  {journal} {\bibinfo
  {journal} {Nature Communications}\ }\textbf {\bibinfo {volume} {7}},\
  \bibinfo {pages} {10793} (\bibinfo {year} {2016})},\ \bibinfo {note}
  {article}\BibitemShut {NoStop}%
\bibitem [{\citenamefont {Alonso-Mora}\ \emph {et~al.}(2017)\citenamefont
  {Alonso-Mora}, \citenamefont {Samaranayake}, \citenamefont {Wallar},
  \citenamefont {Frazzoli},\ and\ \citenamefont {Rus}}]{AlonsoMora2017}%
  \BibitemOpen
  \bibfield  {author} {\bibinfo {author} {\bibfnamefont {Javier}\ \bibnamefont
  {Alonso-Mora}}, \bibinfo {author} {\bibfnamefont {Samitha}\ \bibnamefont
  {Samaranayake}}, \bibinfo {author} {\bibfnamefont {Alex}\ \bibnamefont
  {Wallar}}, \bibinfo {author} {\bibfnamefont {Emilio}\ \bibnamefont
  {Frazzoli}}, \ and\ \bibinfo {author} {\bibfnamefont {Daniela}\ \bibnamefont
  {Rus}},\ }\bibfield  {title} {\enquote {\bibinfo {title} {On-demand
  high-capacity ride-sharing via dynamic trip-vehicle assignment},}\ }\href
  {\doibase 10.1073/pnas.1611675114} {\bibfield  {journal} {\bibinfo  {journal}
  {Proceedings of the National Academy of Sciences}\ }\textbf {\bibinfo
  {volume} {114}},\ \bibinfo {pages} {462--467} (\bibinfo {year}
  {2017})}\BibitemShut {NoStop}%
\bibitem [{\citenamefont {Lim}\ \emph {et~al.}(2011)\citenamefont {Lim},
  \citenamefont {Balakrishnan}, \citenamefont {Gifford}, \citenamefont
  {Madden},\ and\ \citenamefont {Rus}}]{Lim2011}%
  \BibitemOpen
  \bibfield  {author} {\bibinfo {author} {\bibfnamefont {Sejoon}\ \bibnamefont
  {Lim}}, \bibinfo {author} {\bibfnamefont {Hari}\ \bibnamefont
  {Balakrishnan}}, \bibinfo {author} {\bibfnamefont {David}\ \bibnamefont
  {Gifford}}, \bibinfo {author} {\bibfnamefont {Samuel}\ \bibnamefont
  {Madden}}, \ and\ \bibinfo {author} {\bibfnamefont {Daniela}\ \bibnamefont
  {Rus}},\ }\bibfield  {title} {\enquote {\bibinfo {title} {Stochastic motion
  planning and applications to traffic},}\ }\href {\doibase
  10.1177/0278364910386259} {\bibfield  {journal} {\bibinfo  {journal} {The
  International Journal of Robotics Research}\ }\textbf {\bibinfo {volume}
  {30}},\ \bibinfo {pages} {699--712} (\bibinfo {year} {2011})}\BibitemShut
  {NoStop}%
\bibitem [{\citenamefont {Thrun}\ \emph {et~al.}(2006)\citenamefont {Thrun}
  \emph {et~al.}}]{Thrun2006}%
  \BibitemOpen
  \bibfield  {author} {\bibinfo {author} {\bibfnamefont {Sebastian}\
  \bibnamefont {Thrun}} \emph {et~al.},\ }\bibfield  {title} {\enquote
  {\bibinfo {title} {Stanley: The robot that won the darpa grand challenge},}\
  }\href {\doibase 10.1002/rob.20147} {\bibfield  {journal} {\bibinfo
  {journal} {Journal of Field Robotics}\ }\textbf {\bibinfo {volume} {23}},\
  \bibinfo {pages} {661--692} (\bibinfo {year} {2006})}\BibitemShut {NoStop}%
\bibitem [{\citenamefont {Fagnant}\ and\ \citenamefont
  {Kockelman}(2015)}]{Fagnant2015}%
  \BibitemOpen
  \bibfield  {author} {\bibinfo {author} {\bibfnamefont {Daniel~J.}\
  \bibnamefont {Fagnant}}\ and\ \bibinfo {author} {\bibfnamefont {Kara}\
  \bibnamefont {Kockelman}},\ }\bibfield  {title} {\enquote {\bibinfo {title}
  {Preparing a nation for autonomous vehicles: opportunities, barriers and
  policy recommendations},}\ }\href {\doibase
  https://doi.org/10.1016/j.tra.2015.04.003} {\bibfield  {journal} {\bibinfo
  {journal} {Transportation Research Part A: Policy and Practice}\ }\textbf
  {\bibinfo {volume} {77}},\ \bibinfo {pages} {167 -- 181} (\bibinfo {year}
  {2015})}\BibitemShut {NoStop}%
\bibitem [{\citenamefont {{Macfarlane}}(2019)}]{Macfarlane2019}%
  \BibitemOpen
  \bibfield  {author} {\bibinfo {author} {\bibfnamefont {J.}~\bibnamefont
  {{Macfarlane}}},\ }\bibfield  {title} {\enquote {\bibinfo {title} {When apps
  rule the road: The proliferation of navigation apps is causing traffic chaos.
  it's time to restore order},}\ }\href {\doibase 10.1109/MSPEC.2019.8847586}
  {\bibfield  {journal} {\bibinfo  {journal} {IEEE Spectrum}\ }\textbf
  {\bibinfo {volume} {56}},\ \bibinfo {pages} {22--27} (\bibinfo {year}
  {2019})}\BibitemShut {NoStop}%
\bibitem [{\citenamefont {Wu}\ \emph {et~al.}(2017)\citenamefont {Wu},
  \citenamefont {Kreidieh}, \citenamefont {Vinitsky},\ and\ \citenamefont
  {Bayen}}]{pmlr-v78-wu17a}%
  \BibitemOpen
  \bibfield  {author} {\bibinfo {author} {\bibfnamefont {Cathy}\ \bibnamefont
  {Wu}}, \bibinfo {author} {\bibfnamefont {Aboudy}\ \bibnamefont {Kreidieh}},
  \bibinfo {author} {\bibfnamefont {Eugene}\ \bibnamefont {Vinitsky}}, \ and\
  \bibinfo {author} {\bibfnamefont {Alexandre~M.}\ \bibnamefont {Bayen}},\
  }\bibfield  {title} {\enquote {\bibinfo {title} {Emergent behaviors in
  mixed-autonomy traffic},}\ }in\ \href
  {http://proceedings.mlr.press/v78/wu17a.html} {\emph {\bibinfo {booktitle}
  {Proceedings of the 1st Annual Conference on Robot Learning}}},\ \bibinfo
  {series} {Proceedings of Machine Learning Research}, Vol.~\bibinfo {volume}
  {78},\ \bibinfo {editor} {edited by\ \bibinfo {editor} {\bibfnamefont
  {Sergey}\ \bibnamefont {Levine}}, \bibinfo {editor} {\bibfnamefont {Vincent}\
  \bibnamefont {Vanhoucke}}, \ and\ \bibinfo {editor} {\bibfnamefont {Ken}\
  \bibnamefont {Goldberg}}}\ (\bibinfo  {publisher} {PMLR},\ \bibinfo {year}
  {2017})\ pp.\ \bibinfo {pages} {398--407}\BibitemShut {NoStop}%
\bibitem [{\citenamefont {{Kai Nagel}}\ and\ \citenamefont {{Michael
  Schreckenberg}}(1992)}]{Nagel1992}%
  \BibitemOpen
  \bibfield  {author} {\bibinfo {author} {\bibnamefont {{Kai Nagel}}}\ and\
  \bibinfo {author} {\bibnamefont {{Michael Schreckenberg}}},\ }\bibfield
  {title} {\enquote {\bibinfo {title} {A cellular automaton model for freeway
  traffic},}\ }\href {\doibase 10.1051/jp1:1992277} {\bibfield  {journal}
  {\bibinfo  {journal} {J. Phys. I France}\ }\textbf {\bibinfo {volume} {2}},\
  \bibinfo {pages} {2221--2229} (\bibinfo {year} {1992})}\BibitemShut {NoStop}%
\bibitem [{\citenamefont {Bar-Gera}(1999)}]{Bar-Gera1999}%
  \BibitemOpen
  \bibfield  {author} {\bibinfo {author} {\bibfnamefont {Hillel}\ \bibnamefont
  {Bar-Gera}},\ }\emph {\bibinfo {title} {Origin-based Algorithms for
  Transportation Network Modeling}},\ \href@noop {} {Ph.D. thesis},\ \bibinfo
  {address} {Chicago, IL, USA} (\bibinfo {year} {1999}),\ \bibinfo {note}
  {aAI9954880}\BibitemShut {NoStop}%
\bibitem [{\citenamefont {Roughgarden}(2005)}]{Roughgarden2005}%
  \BibitemOpen
  \bibfield  {author} {\bibinfo {author} {\bibfnamefont {Tim}\ \bibnamefont
  {Roughgarden}},\ }\href@noop {} {\emph {\bibinfo {title} {Selfish Routing and
  the Price of Anarchy}}}\ (\bibinfo  {publisher} {The MIT Press},\ \bibinfo
  {year} {2005})\BibitemShut {NoStop}%
\bibitem [{\citenamefont {Szeto}\ and\ \citenamefont {Wong}(2012)}]{Szeto2012}%
  \BibitemOpen
  \bibfield  {author} {\bibinfo {author} {\bibfnamefont {W.~Y.}\ \bibnamefont
  {Szeto}}\ and\ \bibinfo {author} {\bibfnamefont {S.~C.}\ \bibnamefont
  {Wong}},\ }\bibfield  {title} {\enquote {\bibinfo {title} {Dynamic traffic
  assignment: model classifications and recent advances in travel choice
  principles},}\ }\href {\doibase 10.2478/s13531-011-0057-y} {\bibfield
  {journal} {\bibinfo  {journal} {Central European Journal of Engineering}\
  }\textbf {\bibinfo {volume} {2}},\ \bibinfo {pages} {1--18} (\bibinfo {year}
  {2012})}\BibitemShut {NoStop}%
\bibitem [{\citenamefont {Friesz}\ \emph {et~al.}(1989)\citenamefont {Friesz},
  \citenamefont {Luque}, \citenamefont {Tobin},\ and\ \citenamefont
  {Wie}}]{Friesz1989}%
  \BibitemOpen
  \bibfield  {author} {\bibinfo {author} {\bibfnamefont {Terry~L.}\
  \bibnamefont {Friesz}}, \bibinfo {author} {\bibfnamefont {Javier}\
  \bibnamefont {Luque}}, \bibinfo {author} {\bibfnamefont {Roger~L.}\
  \bibnamefont {Tobin}}, \ and\ \bibinfo {author} {\bibfnamefont {Byung-Wook}\
  \bibnamefont {Wie}},\ }\bibfield  {title} {\enquote {\bibinfo {title}
  {Dynamic network traffic assignment considered as a continuous time optimal
  control problem},}\ }\href {http://www.jstor.org/stable/171471} {\bibfield
  {journal} {\bibinfo  {journal} {Operations Research}\ }\textbf {\bibinfo
  {volume} {37}},\ \bibinfo {pages} {893--901} (\bibinfo {year}
  {1989})}\BibitemShut {NoStop}%
\bibitem [{\citenamefont {Ben-Elia}\ and\ \citenamefont
  {Shiftan}(2010)}]{Ben-Elia2010}%
  \BibitemOpen
  \bibfield  {author} {\bibinfo {author} {\bibfnamefont {Eran}\ \bibnamefont
  {Ben-Elia}}\ and\ \bibinfo {author} {\bibfnamefont {Yoram}\ \bibnamefont
  {Shiftan}},\ }\bibfield  {title} {\enquote {\bibinfo {title} {Which road do i
  take? a learning-based model of route-choice behavior with real-time
  information},}\ }\href {\doibase https://doi.org/10.1016/j.tra.2010.01.007}
  {\bibfield  {journal} {\bibinfo  {journal} {Transportation Research Part A:
  Policy and Practice}\ }\textbf {\bibinfo {volume} {44}},\ \bibinfo {pages}
  {249 -- 264} (\bibinfo {year} {2010})}\BibitemShut {NoStop}%
\bibitem [{\citenamefont {Ben-Elia}\ \emph {et~al.}(2013)\citenamefont
  {Ben-Elia}, \citenamefont {Pace}, \citenamefont {Bifulco},\ and\
  \citenamefont {Shiftan}}]{Ben-Elia2013}%
  \BibitemOpen
  \bibfield  {author} {\bibinfo {author} {\bibfnamefont {Eran}\ \bibnamefont
  {Ben-Elia}}, \bibinfo {author} {\bibfnamefont {Roberta~Di}\ \bibnamefont
  {Pace}}, \bibinfo {author} {\bibfnamefont {Gennaro~N.}\ \bibnamefont
  {Bifulco}}, \ and\ \bibinfo {author} {\bibfnamefont {Yoram}\ \bibnamefont
  {Shiftan}},\ }\bibfield  {title} {\enquote {\bibinfo {title} {The impact of
  travel information's accuracy on route-choice},}\ }\href {\doibase
  https://doi.org/10.1016/j.trc.2012.07.001} {\bibfield  {journal} {\bibinfo
  {journal} {Transportation Research Part C: Emerging Technologies}\ }\textbf
  {\bibinfo {volume} {26}},\ \bibinfo {pages} {146 -- 159} (\bibinfo {year}
  {2013})}\BibitemShut {NoStop}%
\bibitem [{\citenamefont {Ben-Akiva}\ \emph {et~al.}(1991)\citenamefont
  {Ben-Akiva}, \citenamefont {Palma},\ and\ \citenamefont
  {Isam}}]{BenAkiva1991}%
  \BibitemOpen
  \bibfield  {author} {\bibinfo {author} {\bibfnamefont {Moshe}\ \bibnamefont
  {Ben-Akiva}}, \bibinfo {author} {\bibfnamefont {Andre~De}\ \bibnamefont
  {Palma}}, \ and\ \bibinfo {author} {\bibfnamefont {Kaysi}\ \bibnamefont
  {Isam}},\ }\bibfield  {title} {\enquote {\bibinfo {title} {Dynamic network
  models and driver information systems},}\ }\href {\doibase
  https://doi.org/10.1016/0191-2607(91)90142-D} {\bibfield  {journal} {\bibinfo
   {journal} {Transportation Research Part A: General}\ }\textbf {\bibinfo
  {volume} {25}},\ \bibinfo {pages} {251 -- 266} (\bibinfo {year}
  {1991})}\BibitemShut {NoStop}%
\bibitem [{\citenamefont {Kuwahara}\ and\ \citenamefont
  {Akamatsu}(1997)}]{Kuwahara1997}%
  \BibitemOpen
  \bibfield  {author} {\bibinfo {author} {\bibfnamefont {Masao}\ \bibnamefont
  {Kuwahara}}\ and\ \bibinfo {author} {\bibfnamefont {Takashi}\ \bibnamefont
  {Akamatsu}},\ }\bibfield  {title} {\enquote {\bibinfo {title} {Decomposition
  of the reactive dynamic assignments with queues for a many-to-many
  origin-destination pattern},}\ }\href {\doibase
  https://doi.org/10.1016/S0191-2615(96)00020-3} {\bibfield  {journal}
  {\bibinfo  {journal} {Transportation Research Part B: Methodological}\
  }\textbf {\bibinfo {volume} {31}},\ \bibinfo {pages} {1 -- 10} (\bibinfo
  {year} {1997})}\BibitemShut {NoStop}%
\bibitem [{\citenamefont {Pel}\ \emph {et~al.}(2009)\citenamefont {Pel},
  \citenamefont {Bliemer},\ and\ \citenamefont {Hoogendoorn}}]{Pel2009}%
  \BibitemOpen
  \bibfield  {author} {\bibinfo {author} {\bibfnamefont {Adam~J.}\ \bibnamefont
  {Pel}}, \bibinfo {author} {\bibfnamefont {Michiel C.~J.}\ \bibnamefont
  {Bliemer}}, \ and\ \bibinfo {author} {\bibfnamefont {Serge~P.}\ \bibnamefont
  {Hoogendoorn}},\ }\bibfield  {title} {\enquote {\bibinfo {title} {Hybrid
  route choice modeling in dynamic traffic assignment},}\ }\href {\doibase
  10.3141/2091-11} {\bibfield  {journal} {\bibinfo  {journal} {Transportation
  Research Record}\ }\textbf {\bibinfo {volume} {2091}},\ \bibinfo {pages}
  {100--107} (\bibinfo {year} {2009})}\BibitemShut {NoStop}%
\bibitem [{ERP(2019)}]{ERP_Singapore}%
  \BibitemOpen
  \href@noop {} {\enquote {\bibinfo {title} {Electronic road pricing},}\
  }\bibinfo {howpublished}
  {\url{https://web.archive.org/web/20110605101108/http://www.lta.gov.sg/motoring_matters/index_motoring_erp.htm}}
  (\bibinfo {year} {2019})\BibitemShut {NoStop}%
\bibitem [{\citenamefont {Barak}(2019)}]{Barak2019}%
  \BibitemOpen
  \bibfield  {author} {\bibinfo {author} {\bibfnamefont {Naama}\ \bibnamefont
  {Barak}},\ }\href@noop {} {\enquote {\bibinfo {title} {Israel tries battling
  traffic jams with cash handouts},}\ }\bibinfo {howpublished} {ISRAEL21c,
  \url{https://www.israel21c.org/israel-tries-battling-traffic-jams-with-cash-handouts/}}
  (\bibinfo {year} {2019})\BibitemShut {NoStop}%
\bibitem [{\citenamefont {Greenshields}(1936)}]{Greenshields1936}%
  \BibitemOpen
  \bibfield  {author} {\bibinfo {author} {\bibfnamefont {B.~D.}\ \bibnamefont
  {Greenshields}},\ }\bibfield  {title} {\enquote {\bibinfo {title} {Studying
  traffic capacity by new methods},}\ }\href {\doibase 10.1037/h0063672}
  {\bibfield  {journal} {\bibinfo  {journal} {J.~Appl.~Psychol.}\ }\textbf
  {\bibinfo {volume} {20}},\ \bibinfo {pages} {353--358} (\bibinfo {year}
  {1936})}\BibitemShut {NoStop}%
\bibitem [{Li2()}]{Li2020sup}%
  \BibitemOpen
  \href@noop {} {}\bibinfo {howpublished} {See Supplemental Material for
  details, which includes Refs. \cite{Wachter2006, Dunning2017}.}\BibitemShut
  {Stop}%
\bibitem [{\citenamefont {Chen}\ and\ \citenamefont {Kempe}(2008)}]{Chen2008}%
  \BibitemOpen
  \bibfield  {author} {\bibinfo {author} {\bibfnamefont {Po-An}\ \bibnamefont
  {Chen}}\ and\ \bibinfo {author} {\bibfnamefont {David}\ \bibnamefont
  {Kempe}},\ }\bibfield  {title} {\enquote {\bibinfo {title} {Altruism,
  selfishness, and spite in traffic routing},}\ }in\ \href {\doibase
  10.1145/1386790.1386816} {\emph {\bibinfo {booktitle} {Proceedings of the 9th
  {ACM} conference on Electronic commerce}}}\ (\bibinfo  {publisher} {{ACM}
  Press},\ \bibinfo {year} {2008})\ pp.\ \bibinfo {pages}
  {140--149}\BibitemShut {NoStop}%
\bibitem [{\citenamefont {Zeng}\ \emph {et~al.}(2019)\citenamefont {Zeng},
  \citenamefont {Li}, \citenamefont {Guo}, \citenamefont {Gao}, \citenamefont
  {Gao}, \citenamefont {Stanley},\ and\ \citenamefont {Havlin}}]{Zeng2019}%
  \BibitemOpen
  \bibfield  {author} {\bibinfo {author} {\bibfnamefont {Guanwen}\ \bibnamefont
  {Zeng}}, \bibinfo {author} {\bibfnamefont {Daqing}\ \bibnamefont {Li}},
  \bibinfo {author} {\bibfnamefont {Shengmin}\ \bibnamefont {Guo}}, \bibinfo
  {author} {\bibfnamefont {Liang}\ \bibnamefont {Gao}}, \bibinfo {author}
  {\bibfnamefont {Ziyou}\ \bibnamefont {Gao}}, \bibinfo {author} {\bibfnamefont
  {H.~Eugene}\ \bibnamefont {Stanley}}, \ and\ \bibinfo {author} {\bibfnamefont
  {Shlomo}\ \bibnamefont {Havlin}},\ }\bibfield  {title} {\enquote {\bibinfo
  {title} {Switch between critical percolation modes in city traffic
  dynamics},}\ }\href {\doibase 10.1073/pnas.1801545116} {\bibfield  {journal}
  {\bibinfo  {journal} {Proceedings of the National Academy of Sciences}\
  }\textbf {\bibinfo {volume} {116}},\ \bibinfo {pages} {23--28} (\bibinfo
  {year} {2019})}\BibitemShut {NoStop}%
\bibitem [{\citenamefont {{OpenStreetMap contributors}}(2017)}]{OpenStreetMap}%
  \BibitemOpen
  \bibfield  {author} {\bibinfo {author} {\bibnamefont {{OpenStreetMap
  contributors}}},\ }\href@noop {} {\enquote {\bibinfo {title} {{Planet dump
  retrieved from https://planet.osm.org}},}\ }\bibinfo {howpublished}
  {\url{https://www.openstreetmap.org}} (\bibinfo {year} {2017})\BibitemShut
  {NoStop}%
\bibitem [{\citenamefont {Karduni}\ \emph {et~al.}(2016)\citenamefont
  {Karduni}, \citenamefont {Kermanshah},\ and\ \citenamefont
  {Derrible}}]{Karduni2016}%
  \BibitemOpen
  \bibfield  {author} {\bibinfo {author} {\bibfnamefont {Alireza}\ \bibnamefont
  {Karduni}}, \bibinfo {author} {\bibfnamefont {Amirhassan}\ \bibnamefont
  {Kermanshah}}, \ and\ \bibinfo {author} {\bibfnamefont {Sybil}\ \bibnamefont
  {Derrible}},\ }\bibfield  {title} {\enquote {\bibinfo {title} {A protocol to
  convert spatial polyline data to network formats and applications to world
  urban road networks},}\ }\href {https://doi.org/10.1038/sdata.2016.46}
  {\bibfield  {journal} {\bibinfo  {journal} {Scientific Data}\ }\textbf
  {\bibinfo {volume} {3}},\ \bibinfo {pages} {160046 EP --} (\bibinfo {year}
  {2016})},\ \bibinfo {note} {data Descriptor}\BibitemShut {NoStop}%
\bibitem [{\citenamefont {Braess}(1968)}]{Braess1968}%
  \BibitemOpen
  \bibfield  {author} {\bibinfo {author} {\bibfnamefont {D.}~\bibnamefont
  {Braess}},\ }\bibfield  {title} {\enquote {\bibinfo {title} {{\"U}ber ein
  paradoxon aus der verkehrsplanung},}\ }\href {\doibase 10.1007/BF01918335}
  {\bibfield  {journal} {\bibinfo  {journal} {Unternehmensforschung}\ }\textbf
  {\bibinfo {volume} {12}},\ \bibinfo {pages} {258--268} (\bibinfo {year}
  {1968})}\BibitemShut {NoStop}%
\bibitem [{\citenamefont {Cohen}\ and\ \citenamefont
  {Horowitz}(1991)}]{Cohen1991}%
  \BibitemOpen
  \bibfield  {author} {\bibinfo {author} {\bibfnamefont {Joel~E.}\ \bibnamefont
  {Cohen}}\ and\ \bibinfo {author} {\bibfnamefont {Paul}\ \bibnamefont
  {Horowitz}},\ }\bibfield  {title} {\enquote {\bibinfo {title} {Paradoxical
  behaviour of mechanical and electrical networks},}\ }\href {\doibase
  10.1038/352699a0} {\bibfield  {journal} {\bibinfo  {journal} {Nature}\
  }\textbf {\bibinfo {volume} {352}},\ \bibinfo {pages} {699--701} (\bibinfo
  {year} {1991})}\BibitemShut {NoStop}%
\bibitem [{\citenamefont {Witthaut}\ and\ \citenamefont
  {Timme}(2012)}]{Witthaut2012}%
  \BibitemOpen
  \bibfield  {author} {\bibinfo {author} {\bibfnamefont {Dirk}\ \bibnamefont
  {Witthaut}}\ and\ \bibinfo {author} {\bibfnamefont {Marc}\ \bibnamefont
  {Timme}},\ }\bibfield  {title} {\enquote {\bibinfo {title} {Braess's paradox
  in oscillator networks, desynchronization and power outage},}\ }\href
  {\doibase 10.1088/1367-2630/14/8/083036} {\bibfield  {journal} {\bibinfo
  {journal} {New Journal of Physics}\ }\textbf {\bibinfo {volume} {14}},\
  \bibinfo {pages} {083036} (\bibinfo {year} {2012})}\BibitemShut {NoStop}%
\bibitem [{\citenamefont {Donovan}(2018)}]{Donovan2018}%
  \BibitemOpen
  \bibfield  {author} {\bibinfo {author} {\bibfnamefont {Graham~M.}\
  \bibnamefont {Donovan}},\ }\bibfield  {title} {\enquote {\bibinfo {title}
  {Biological version of braess' paradox arising from perturbed homeostasis},}\
  }\href {\doibase 10.1103/PhysRevE.98.062406} {\bibfield  {journal} {\bibinfo
  {journal} {Phys. Rev. E}\ }\textbf {\bibinfo {volume} {98}},\ \bibinfo
  {pages} {062406} (\bibinfo {year} {2018})}\BibitemShut {NoStop}%
\bibitem [{\citenamefont {Chernousko}\ and\ \citenamefont
  {Lyubushin}(1982)}]{Chernousko1982}%
  \BibitemOpen
  \bibfield  {author} {\bibinfo {author} {\bibfnamefont {F.~L.}\ \bibnamefont
  {Chernousko}}\ and\ \bibinfo {author} {\bibfnamefont {A.~A.}\ \bibnamefont
  {Lyubushin}},\ }\bibfield  {title} {\enquote {\bibinfo {title} {Method of
  successive approximations for solution of optimal control problems},}\ }\href
  {\doibase 10.1002/oca.4660030201} {\bibfield  {journal} {\bibinfo  {journal}
  {Optimal Control Applications and Methods}\ }\textbf {\bibinfo {volume}
  {3}},\ \bibinfo {pages} {101--114} (\bibinfo {year} {1982})}\BibitemShut
  {NoStop}%
\bibitem [{\citenamefont {Lokhov}\ and\ \citenamefont
  {Saad}(2017)}]{Lokhov2017}%
  \BibitemOpen
  \bibfield  {author} {\bibinfo {author} {\bibfnamefont {Andrey~Y.}\
  \bibnamefont {Lokhov}}\ and\ \bibinfo {author} {\bibfnamefont {David}\
  \bibnamefont {Saad}},\ }\bibfield  {title} {\enquote {\bibinfo {title}
  {Optimal deployment of resources for maximizing impact in spreading
  processes},}\ }\href {\doibase 10.1073/pnas.1614694114} {\bibfield  {journal}
  {\bibinfo  {journal} {Proceedings of the National Academy of Sciences}\
  }\textbf {\bibinfo {volume} {114}},\ \bibinfo {pages} {E8138--E8146}
  (\bibinfo {year} {2017})}\BibitemShut {NoStop}%
\bibitem [{\citenamefont {Li}\ \emph {et~al.}(2018)\citenamefont {Li},
  \citenamefont {Chen}, \citenamefont {Tai},\ and\ \citenamefont
  {E}}]{QianxiaoLi2017}%
  \BibitemOpen
  \bibfield  {author} {\bibinfo {author} {\bibfnamefont {Qianxiao}\
  \bibnamefont {Li}}, \bibinfo {author} {\bibfnamefont {Long}\ \bibnamefont
  {Chen}}, \bibinfo {author} {\bibfnamefont {Cheng}\ \bibnamefont {Tai}}, \
  and\ \bibinfo {author} {\bibfnamefont {Weinan}\ \bibnamefont {E}},\
  }\bibfield  {title} {\enquote {\bibinfo {title} {Maximum principle based
  algorithms for deep learning},}\ }\href
  {http://jmlr.org/papers/v18/17-653.html} {\bibfield  {journal} {\bibinfo
  {journal} {Journal of Machine Learning Research}\ }\textbf {\bibinfo {volume}
  {18}},\ \bibinfo {pages} {1--29} (\bibinfo {year} {2018})}\BibitemShut
  {NoStop}%
\bibitem [{\citenamefont {W{\"a}chter}\ and\ \citenamefont
  {Biegler}(2006)}]{Wachter2006}%
  \BibitemOpen
  \bibfield  {author} {\bibinfo {author} {\bibfnamefont {Andreas}\ \bibnamefont
  {W{\"a}chter}}\ and\ \bibinfo {author} {\bibfnamefont {Lorenz~T.}\
  \bibnamefont {Biegler}},\ }\bibfield  {title} {\enquote {\bibinfo {title} {On
  the implementation of an interior-point filter line-search algorithm for
  large-scale nonlinear programming},}\ }\href {\doibase
  10.1007/s10107-004-0559-y} {\bibfield  {journal} {\bibinfo  {journal}
  {Mathematical Programming}\ }\textbf {\bibinfo {volume} {106}},\ \bibinfo
  {pages} {25--57} (\bibinfo {year} {2006})}\BibitemShut {NoStop}%
\bibitem [{\citenamefont {Dunning}\ \emph {et~al.}(2017)\citenamefont
  {Dunning}, \citenamefont {Huchette},\ and\ \citenamefont
  {Lubin}}]{Dunning2017}%
  \BibitemOpen
  \bibfield  {author} {\bibinfo {author} {\bibfnamefont {Iain}\ \bibnamefont
  {Dunning}}, \bibinfo {author} {\bibfnamefont {Joey}\ \bibnamefont
  {Huchette}}, \ and\ \bibinfo {author} {\bibfnamefont {Miles}\ \bibnamefont
  {Lubin}},\ }\bibfield  {title} {\enquote {\bibinfo {title} {Jump: A modeling
  language for mathematical optimization},}\ }\href {\doibase
  10.1137/15M1020575} {\bibfield  {journal} {\bibinfo  {journal} {SIAM Review}\
  }\textbf {\bibinfo {volume} {59}},\ \bibinfo {pages} {295--320} (\bibinfo
  {year} {2017})}\BibitemShut {NoStop}%
\end{thebibliography}%


\begin{thebibliography}{6}%
\makeatletter
\providecommand \@ifxundefined [1]{%
 \@ifx{#1\undefined}
}%
\providecommand \@ifnum [1]{%
 \ifnum #1\expandafter \@firstoftwo
 \else \expandafter \@secondoftwo
 \fi
}%
\providecommand \@ifx [1]{%
 \ifx #1\expandafter \@firstoftwo
 \else \expandafter \@secondoftwo
 \fi
}%
\providecommand \natexlab [1]{#1}%
\providecommand \enquote  [1]{``#1''}%
\providecommand \bibnamefont  [1]{#1}%
\providecommand \bibfnamefont [1]{#1}%
\providecommand \citenamefont [1]{#1}%
\providecommand \href@noop [0]{\@secondoftwo}%
\providecommand \href [0]{\begingroup \@sanitize@url \@href}%
\providecommand \@href[1]{\@@startlink{#1}\@@href}%
\providecommand \@@href[1]{\endgroup#1\@@endlink}%
\providecommand \@sanitize@url [0]{\catcode `\\12\catcode `\$12\catcode
  `\&12\catcode `\#12\catcode `\^12\catcode `\_12\catcode `\%12\relax}%
\providecommand \@@startlink[1]{}%
\providecommand \@@endlink[0]{}%
\providecommand \url  [0]{\begingroup\@sanitize@url \@url }%
\providecommand \@url [1]{\endgroup\@href {#1}{\urlprefix }}%
\providecommand \urlprefix  [0]{URL }%
\providecommand \Eprint [0]{\href }%
\providecommand \doibase [0]{http://dx.doi.org/}%
\providecommand \selectlanguage [0]{\@gobble}%
\providecommand \bibinfo  [0]{\@secondoftwo}%
\providecommand \bibfield  [0]{\@secondoftwo}%
\providecommand \translation [1]{[#1]}%
\providecommand \BibitemOpen [0]{}%
\providecommand \bibitemStop [0]{}%
\providecommand \bibitemNoStop [0]{.\EOS\space}%
\providecommand \EOS [0]{\spacefactor3000\relax}%
\providecommand \BibitemShut  [1]{\csname bibitem#1\endcsname}%
\let\auto@bib@innerbib\@empty
\bibitem [{\citenamefont {Greenshields}(1936)}]{Greenshields1936}%
  \BibitemOpen
  \bibfield  {author} {\bibinfo {author} {\bibfnamefont {B.~D.}\ \bibnamefont
  {Greenshields}},\ }\bibfield  {title} {\enquote {\bibinfo {title} {Studying
  traffic capacity by new methods},}\ }\href {\doibase 10.1037/h0063672}
  {\bibfield  {journal} {\bibinfo  {journal} {J.~Appl.~Psychol.}\ }\textbf
  {\bibinfo {volume} {20}},\ \bibinfo {pages} {353--358} (\bibinfo {year}
  {1936})}\BibitemShut {NoStop}%
\bibitem [{\citenamefont {Chernousko}\ and\ \citenamefont
  {Lyubushin}(1982)}]{Chernousko1982}%
  \BibitemOpen
  \bibfield  {author} {\bibinfo {author} {\bibfnamefont {F.~L.}\ \bibnamefont
  {Chernousko}}\ and\ \bibinfo {author} {\bibfnamefont {A.~A.}\ \bibnamefont
  {Lyubushin}},\ }\bibfield  {title} {\enquote {\bibinfo {title} {Method of
  successive approximations for solution of optimal control problems},}\ }\href
  {\doibase 10.1002/oca.4660030201} {\bibfield  {journal} {\bibinfo  {journal}
  {Optimal Control Applications and Methods}\ }\textbf {\bibinfo {volume}
  {3}},\ \bibinfo {pages} {101--114} (\bibinfo {year} {1982})}\BibitemShut
  {NoStop}%
\bibitem [{\citenamefont {Lokhov}\ and\ \citenamefont
  {Saad}(2017)}]{Lokhov2017}%
  \BibitemOpen
  \bibfield  {author} {\bibinfo {author} {\bibfnamefont {Andrey~Y.}\
  \bibnamefont {Lokhov}}\ and\ \bibinfo {author} {\bibfnamefont {David}\
  \bibnamefont {Saad}},\ }\bibfield  {title} {\enquote {\bibinfo {title}
  {Optimal deployment of resources for maximizing impact in spreading
  processes},}\ }\href {\doibase 10.1073/pnas.1614694114} {\bibfield  {journal}
  {\bibinfo  {journal} {Proceedings of the National Academy of Sciences}\
  }\textbf {\bibinfo {volume} {114}},\ \bibinfo {pages} {E8138--E8146}
  (\bibinfo {year} {2017})}\BibitemShut {NoStop}%
\bibitem [{\citenamefont {Li}\ \emph {et~al.}(2018)\citenamefont {Li},
  \citenamefont {Chen}, \citenamefont {Tai},\ and\ \citenamefont
  {E}}]{QianxiaoLi2017}%
  \BibitemOpen
  \bibfield  {author} {\bibinfo {author} {\bibfnamefont {Qianxiao}\
  \bibnamefont {Li}}, \bibinfo {author} {\bibfnamefont {Long}\ \bibnamefont
  {Chen}}, \bibinfo {author} {\bibfnamefont {Cheng}\ \bibnamefont {Tai}}, \
  and\ \bibinfo {author} {\bibfnamefont {Weinan}\ \bibnamefont {E}},\
  }\bibfield  {title} {\enquote {\bibinfo {title} {Maximum principle based
  algorithms for deep learning},}\ }\href
  {http://jmlr.org/papers/v18/17-653.html} {\bibfield  {journal} {\bibinfo
  {journal} {Journal of Machine Learning Research}\ }\textbf {\bibinfo {volume}
  {18}},\ \bibinfo {pages} {1--29} (\bibinfo {year} {2018})}\BibitemShut
  {NoStop}%
\bibitem [{\citenamefont {W{\"a}chter}\ and\ \citenamefont
  {Biegler}(2006)}]{Wachter2006}%
  \BibitemOpen
  \bibfield  {author} {\bibinfo {author} {\bibfnamefont {Andreas}\ \bibnamefont
  {W{\"a}chter}}\ and\ \bibinfo {author} {\bibfnamefont {Lorenz~T.}\
  \bibnamefont {Biegler}},\ }\bibfield  {title} {\enquote {\bibinfo {title} {On
  the implementation of an interior-point filter line-search algorithm for
  large-scale nonlinear programming},}\ }\href {\doibase
  10.1007/s10107-004-0559-y} {\bibfield  {journal} {\bibinfo  {journal}
  {Mathematical Programming}\ }\textbf {\bibinfo {volume} {106}},\ \bibinfo
  {pages} {25--57} (\bibinfo {year} {2006})}\BibitemShut {NoStop}%
\bibitem [{\citenamefont {Dunning}\ \emph {et~al.}(2017)\citenamefont
  {Dunning}, \citenamefont {Huchette},\ and\ \citenamefont
  {Lubin}}]{Dunning2017}%
  \BibitemOpen
  \bibfield  {author} {\bibinfo {author} {\bibfnamefont {Iain}\ \bibnamefont
  {Dunning}}, \bibinfo {author} {\bibfnamefont {Joey}\ \bibnamefont
  {Huchette}}, \ and\ \bibinfo {author} {\bibfnamefont {Miles}\ \bibnamefont
  {Lubin}},\ }\bibfield  {title} {\enquote {\bibinfo {title} {Jump: A modeling
  language for mathematical optimization},}\ }\href {\doibase
  10.1137/15M1020575} {\bibfield  {journal} {\bibinfo  {journal} {SIAM Review}\
  }\textbf {\bibinfo {volume} {59}},\ \bibinfo {pages} {295--320} (\bibinfo
  {year} {2017})}\BibitemShut {NoStop}%
\end{thebibliography}%

\end{document}


\title{Reducing Urban Traffic Congestion Due To Localized Routing Decisions\\Supplemental Material}
		
	\author{Bo Li}
	\selectlanguage{english}%
	\affiliation{Non-linearity and Complexity Research Group, Aston University, Birmingham, B4 7ET, United Kingdom}
	
	\author{David Saad}
	\selectlanguage{english}%
	\affiliation{Non-linearity and Complexity Research Group, Aston University, Birmingham, B4 7ET, United Kingdom}

	\author{Andrey Y. Lokhov}
	\selectlanguage{english}%
	\affiliation{Theoretical Division, Los Alamos National Laboratory, Los Alamos, New Mexico 87545, USA}

    \maketitle

\section{The Model}
\label{app:model}

\begin{figure*}
    \includegraphics[scale=0.24]{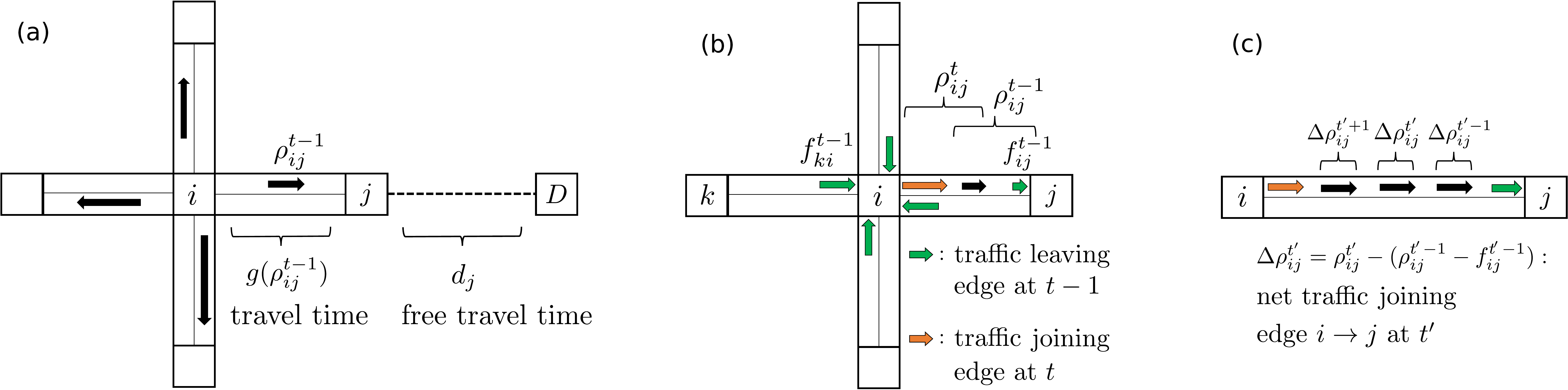}
    \caption{(a) 
    The decision making rule of the drivers who make their own route choice. When the drivers arrive to junction $i$ at time $t-1$, they assess the traffic volumes in nearest-neighbor down-stream roads $\density^{t-1}_{ij}, j\in \partial i$, based on which the corresponding travel are estimated as $g(\density^{t-1}_{ij})$. Since they are not aware of the traffic conditions beyond the nearest neighbour roads, they estimate the remaining time needed to travel from node $j$ to the destination $D$ based on the free travel time, denoted as $d_{j}$. Consequently, their probability for choosing route $i\to j$ is $p^{g,t-1}_{ij} = \exp( -\beta [g(\density^{t-1}_{ij}) + d_j] ) / Z_{g} $, where $Z_{g}$ is a normalization factor. (b) Flow conservation in the discrete dynamics in Eq.~(4). The traffic volume $\density^{t}_{ij}$ comprises (i) the existing users who have not left road $i\to j$ at time $t$ (black arrow); and (ii) newly joined users who selected this road at junction node $i$ at time $t-1$ (the orange arrow). Clearly, we also take account of users who left the road in the previous time step (green arrow). This illustrates the derivation of Eq.~(4). (c) The existing traffic volume on edge $i\to j$ can be decomposed into net traffic volume $\Delta \density^{t'}_{ij}$ increase at different times $t'$; the traffic condition at the time of their entrance (i.e. $\density^{t'}_{ij}$) dictates the probability of travel time $\tau$ needed to complete the journey on this road, denoted as $P(\tau|\density^{t'}_{ij})$. This schematic helps explain the derivation of Eq.~(5).}
    \label{fig:illustrate}
    \end{figure*}

In this section, we provide additional details of the dynamical model proposed in the main text.
Figure~\ref{fig:illustrate} summarizes the decision making rule~Eq.~(1) and dynamical rules~Eq.~(4)-Eq.~(5) defined in the main text.

In the proposed dynamical model, the traffic volume $\density^{t}_{ij}$ is not explicitly limited by an upper bound to obey the road capacity constraint. Instead, a heavily-used road, say edge $i \to j$, takes longer to travel through as reflected by the arrival probability $P(t-t'|\density^{t'}_{ij})$, resulting in a smaller probability of choosing it. This limits the traffic volume $\density^{t}_{ij}$ from growing indefinitely.

The form of the arrival probability $P(t-t'|\density^{t'}_{ij})$ reflects the vehicle flow on edge $i \to j$, where a plausible choice is to match the corresponding mean traveling time $\mathbb{E}_{t \sim P}[t - t']$ to the traveling time $g(\density^{t'}_{ij})$ predicted by a real-world traffic statistics.
For a concrete example, we assume that the arrival time follows a Poisson distribution with the form of $\Poi(t-t'-\tf_{ij}|\lambda^{t'}_{ij}(\density^{t'}_{ij}))$, where $\tf_{ij}$ is the free traveling time on edge $i \to j$, and the parameter $\lambda^{t'}_{ij}$ of the Poisson distribution has the form $\lambda^{t'}_{ij}(\density^{t'}_{ij}) = g(\density^{t'}_{ij}) - \tf_{ij}$, such that the mean traveling time is $\mathbb{E}_{t \sim P}[t - t'] = g(\density^{t'}_{ij})$.
We further model the mean traveling time $g(\density)$ as a function of traffic volume by modifying the Greenshields model~\cite{Greenshields1936} (also see the section below for a brief description) 
    \begin{equation}
        g(\density) = \begin{cases} 
        \tf \frac{1}{1-\density/\densjam}, & \density < \densjam (1 - \epsilon), \\
        \tf / \epsilon, & \density \ge \densjam (1 - \epsilon). \nonumber
        \end{cases}
    \end{equation}
where $\densjam$ is the jam volume defined in Greenshields model, and $\epsilon$ a cutoff parameter to allow for a slow-moving traffic flow even under the regime of severe traffic congestion.

\section{The Greenshields model}
\label{app:greenshields}
The traditional Greenshields model is characterized by a linear relation between traffic speed $u$ and traffic density $k$ (not to be confused with the connectivity $k_i$ of node $i$ in the main text) on a particular road section
    \begin{equation}
    v = v^{\text{free}} \left( 1- \frac{k}{k^{\text{jam}}} \right),
    \end{equation}
where $v^{\text{free}}$ is the free flow speed and $k^{\text{jam}}$ is the jam density. The traffic volume $\density$ is related to the traffic density $k$ and road section length $l$ through $\density = k \cdot l$, and similarly $\densjam = k^{\text{jam}} \cdot l$. Therefore the traveling time on the road is
    \begin{equation}
    t = \frac{l}{v} = \frac{l}{v^{\text{free}}} \frac{1}{1- \frac{\density}{\densjam}},
    \end{equation}
where $l/v^{\text{free}}$ is identified as the free traveling time $\tf$.

\section{Small-world network}
\label{app:smallworld}
The small-world networks considered in the main text are generated according to the following procedure. We first generate a $L_x \times L_y$ square lattice and identify the center node and its nearest neighbors as the city center, and define the city center cluster to be the destination node $D$. We then randomly and independently disconnect each link with a fixed small probability $p_r$ and re-connect one end of the link to a randomly chosen site of the network. For the $21 \times 21$ square lattices considered in the main text, we used $p_r = 0.05$. Non-shortcut links were assumed to be of the same length and were assigned the same parameter values of $\tf = 3$ and $\densjam = 16$. For the shortcut links, we let $\tf$ and $\densjam$ be proportional to their lengths defined as the Euclidean distances between two sites; we also assume that users can drive on the shortcut edge with a higher speed than the non-shortcut edge since the shortcut edges represent high-speed urban roads in this model. 
These speed parameter choices do not seem to change the qualitative behavior we observe. 
   
\section{Birmingham UK road network}
\label{app:birmingham}
The Birmingham road network is extracted from the Open Street Map (OSM) datasets. For computational efficiency, we do not consider residential roads and keep only highways, primary roads and secondary roads and the corresponding ramps as classified by the OSM dataset type. They includes motorways, A roads and B roads in the UK road network. We neglect the M6 motorway in our analysis since it is rarely used in city commute within Birmingham, while its ramps can significantly complicate the network structure. The obtained OSM data comprises the polylines describing the geometric shapes of the roads and the corresponding attributes such as length $l$, maximal speed $v_{\text{max}}$ and number of lanes $n_{\text{lane}}$. 
    
We export these OSM data as in shapefile format and use the GIS F2E software to convert them into networks as described by edge lists and edge attributes. We then determine the free traveling time and jam volume of each edge as $\tf = l / v_{\text{max}}, \densjam = l \cdot n_{\text{lane}}$. The resulting network has a large number of edges, many of which are of very short length. The nodes separating short-length edges are not considered decision-nodes of interest. Therefore, we apply some post-processing operations to the network, primarily by identifying the roundabouts and intersections of roads as decision-making nodes (DMNs), and combine the edges between two adjacent DMNs to a single edge. When we combine two edges $(i, j)$ and $(j, k)$ to form a new edge $(i, k)$, the edge parameters are determined by $\tf_{ik} = \tf_{ij} + \tf_{jk}, \densjam_{ik} = \densjam_{ij} + \densjam_{jk}$. We define $20$ seconds as one time step and round up the number of time steps needed to travel along edge $(i,j)$ to ensure $\tf_{ij}$ is an integer.

\section{Effect of the Parameter $\beta$}
\label{app:vary_beta}
\begin{figure}
        \includegraphics[scale=0.4]{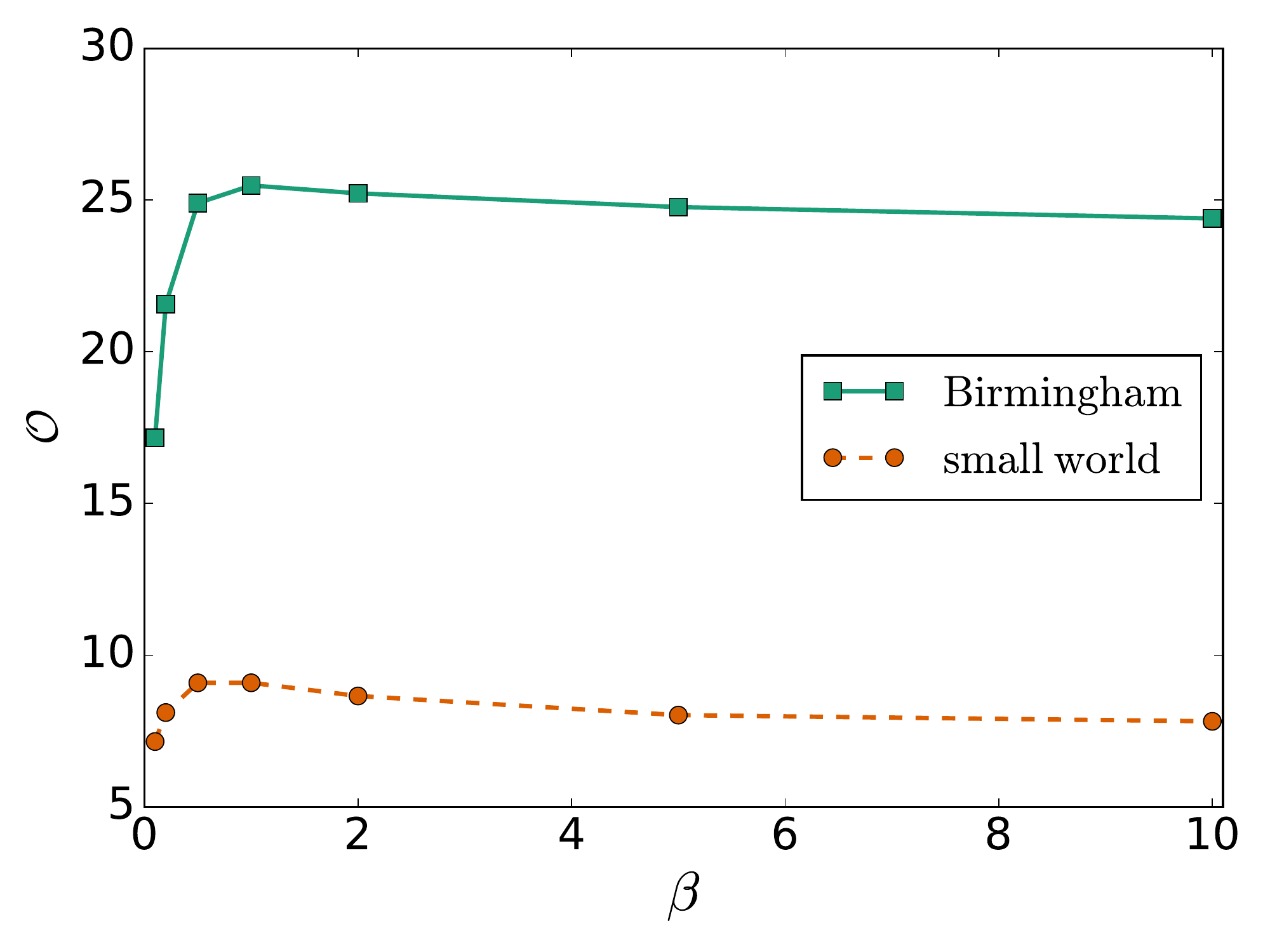}
       \caption{${\cal O}$ vs $\beta$. The other parameters are $T=100$, $n=0$, load level $L=0.2$. The small world network is obtained by rewiring a $21\times 21$ square lattice. }
        \label{fig:vary_beta}
    \end{figure}

In this section, we examine the effect of the parameter $\beta$ in the route choice model defined in the main text. We consider the scenario where the drivers make their own instantaneous decisions, i.e., $n=0$. The relation between the objective function ${\cal O}$ and $\beta$ is depicted in Fig.~\ref{fig:vary_beta}. For $\beta \ll 1$, the users behave like random walkers on the network; typically it takes a long time for them to hit the destination, as indicated by a small value of ${\cal O}$. The objective function ${\cal O}$ reaches the maximum at about $\beta=1$ in both networks considered, and it deteriorates slightly with larger $\beta$. As the system performance is not very sensitive to $\beta$ when $\beta \geq 1$, we fix $\beta=1$ in this study.

\section{System Efficiency as a Function of Load}
\begin{figure}
        \includegraphics[scale=1.6]{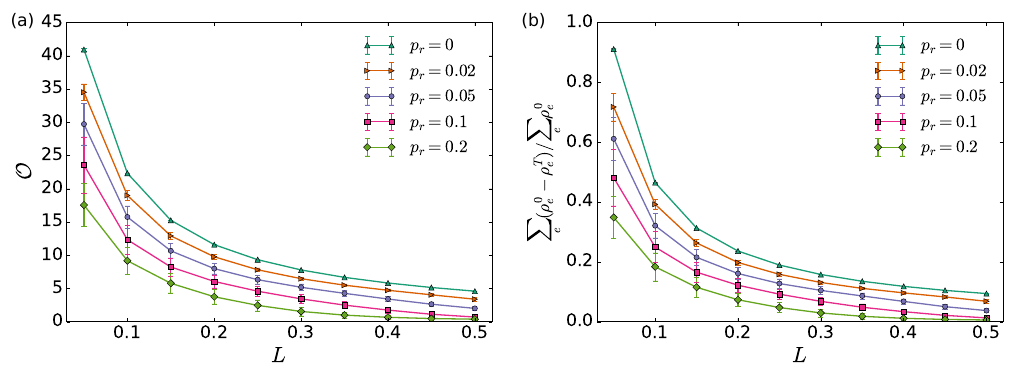}
       \caption{(a) Average arrival-time ahead of $\tmax$ reaching destination as a function of load $L$. (b) Fraction of users reaching destination before $\tmax$ as a function of load. The experiments are performed on $21\! \times\! 21$ small world networks; parameters are $T\!=\!100, n\!=\!0$. For each data point with a specific network rewiring probability $p_r$ and load $L$, we obtained the average behavior of the quantities measured over 10 different network structures and 10 random initial traffic volumes.}
        \label{fig:objf_and_fd_vs_L}
    \end{figure}

In Fig.~\ref{fig:objf_and_fd_vs_L}, we quantify the level of congestion by contrasting the average arrival time to destination ahead of $\tmax$ defined in Eq.~(6) and the fraction of users arriving at destination before $\tmax$, against the load level $L$ on the same small world network ensembles. We observe that as load $L$ increases, both quantities ${\cal O}$ and $\sum_e (\density^{0}_{e} - \density^{\tmax}_{e}) / \sum_{e} \density^{0}_{e}$ decrease rapidly suggesting, unsurprisingly, that a higher usage of the traffic system is prone to lead to congestion in this model. 
The rapid decrease in objective function with increasing demands suggests an increase in travel time which grows faster than linear when more vehicles enter the system. 
Interestingly, as the rewiring probability $p_r$ increases, both ${\cal O}$ and $\sum_e (\density^{0}_{e} - \density^{\tmax}_{e}) / \sum_{e} \density^{0}_{e}$ \emph{drop}, which indicates counter-intuitively, that \emph{additional shortcut edges contribute to a higher level of congestion}. This can be understood by the fact that in the absence of routing advice, drivers are only aware of the local traffic conditions and tend to use the shortcut edges beyond the immediate neighboring junctions, which leads to high latency in the shortcut roads.

\section{Optimization}
\label{app:optimization}

The objective to be maximized is the system efficiency ${\cal O}$ defined in Eq.~(6) of the main text, while obeying the constraints representing the dynamics. In general it is difficult to deal with the non-linear equality constraints imposed by the forward dynamics using the non-linear programming framework. In this work, we adopt the optimal control approach by introducing the Lagrangian  
    \begin{eqnarray}
        {\cal L} &&= {\cal O} + \sum_{t=1}^{\tmax} \sum_{i \neq D} \sum_{j} A_{ij} \mu^{t}_{ij}  \bigg[ -\flux^{t}_{ij} + \sum^{t}_{t'=1}  \big[ \density^{t'}_{ij}\!-\! (\density^{t'-1}_{ij} \!- \!\flux^{t'-1}_{ij}) \big] P(t\!-\!t'|\density^{t'}_{ij}) \!+\! \density^{0}_{ij} P(t|\density^{0}_{ij})  \bigg]  \nonumber \\
        && + \sum_{t=1}^{\tmax} \sum_{i\neq D} \sum_{j} A_{ij} \eta^{t}_{ij} \bigg[ -\density^{t}_{ij} + p^{t-1}_{ij}(\density^{t-1}, w^{t-1})\sum_{ k\in\partial i, k \neq D} \flux^{t-1}_{ki} + (\density^{t-1}_{ij} - \flux^{t-1}_{ij})  \bigg], \label{eq:Lagrangian}
    \end{eqnarray}
where $A_{ij}\in\{0,1\}$ is the element of the adjacency matrix of the underlying graph, taking a value $1$ if a link exists and $0$ otherwise, and $\{ \mu^{t}_{ij}, \eta^{t}_{ij} \}$ are the Lagrange multipliers enforcing the dynamics specified by Eqs.~(4) and~(5), respectively.

The optimal solution can be pursued by extremizing ${\cal L}$ with respect to the state variables $\{ \density^{t}_{ij}, \flux^{t}_{ij} \}$, the Lagrange multipliers $\{ \mu^{t}_{ij}, \eta^{t}_{ij} \}$ and the control variables $\{ w^{t}_{ij} \}$. The stationary point equations $\partial {\cal L} / \partial \eta^{t}_{ij} = 0$ and $\partial {\cal L} / \partial \mu^{t}_{ij} = 0$ return the equations of state, Eqs.~(4) and~(5) propagating forward in time, while $\partial {\cal L} / \partial \density^{t}_{ij} = 0$ and $\partial {\cal L} / \partial \flux^{t}_{ij} = 0$ yield two backward propagation equations of the Lagrange multipliers, signaling the system's deviation from optimum. Equation $\partial {\cal L} / \partial w^{t}_{ij} = 0$ dictates the optimal recommended weights $w^{t}_{ij}$ for the given objective function. The set of stationary point equations are solved by an iterative forward propagation of the state variables, followed by backward propagation of the Lagrange multipliers and an update of the control variables, which resembles the method of successive approximation in optimal control theory~\cite{Chernousko1982}. Similar techniques has been recently introduced for the impact maximization in spreading processes~\cite{Lokhov2017}. 
The control parameters $w^{t}_{ij}$ are gradually updated through gradient ascent as $w^{t}_{ij} \leftarrow w^{t}_{ij} + s \partial {\cal L} / \partial w^{t}_{ij}$ with step size $s$ after a forward-backward propagation sweep, which will gradually drive the system towards a locally optimal state.

More specifically, the task is to maximize the average time ahead of $T$ reaching the destination with respect to $\{ w^{t}_{ij} \}$
    \begin{equation}
        \max_{\bm{w}} {\cal O} (\bm{w}) = \frac{1}{\sum_{e \in \cE} \density^{0}_{e} }\sum_{t=1}^{\tmax} \sum_{j \in \partial D} \flux^{t}_{jD} (\bm{w}) \times (\tmax-t),
    \end{equation}
    such that the following dynamical constraints are fulfilled,
    {\small 
    \begin{eqnarray}
	    \density^{t}_{ij} &=& p^{t-1}_{ij}\sum_{ k\in\partial i, k\neq D} \flux^{t-1}_{ki} + (\density^{t-1}_{ij} - \flux^{t-1}_{ij}), \label{eq:forward1_si} \\
	    \flux^{t}_{ij} &=& \sum^{t}_{t'=1}  \big[ \density^{t'}_{ij}- (\density^{t'-1}_{ij} - \flux^{t'-1}_{ij}) \big] \Poi(t-t'-\tf|\lambda^{t'}_{ij}) \indicator_{t-t' \ge \tf} \nonumber \label{eq:forward2_si} \\
	    &&  + \density^{0}_{ij} \Poi(t-\tf|\lambda^{0}_{ij}) \indicator_{t\ge \tf},
    \end{eqnarray}}
    where $\indicator_{(\cdot)}$ is the indicator function.
    The dynamics satisfies the absorbing boundary conditions $\forall ij\in\cE, i=D$
    \begin{equation}
        \density^{t}_{ij} = 0, \quad \flux^{t}_{ij} = 0, \quad \forall t,
    \end{equation}
    and the initial conditions $\forall ij\in\cE, i \neq D$ 
    \begin{equation}
        \density^{0}_{ij} = \tilde{\density}_{ij}, \quad \flux^{0}_{ij} = \density^{0}_{ij} \Poi(0|\lambda^{0}_{ij}(\density^{0}_{ij})) \delta_{\tf, 0}, \label{eq:init_cond_si}
    \end{equation}
    where $\tilde{\density}_{ij}$ is the predefined initial traffic volume.

    The Lagrangian to be extremized reads
     {\small 
    \begin{eqnarray}
        &&{\cal L}  = \frac{1}{\sum_{e \in \cE} \density^{0}_{e} } \sum_{t=1}^{\tmax} \sum_{j \in \partial D} \flux^{t}_{jD} \cdot (\tmax-t) + \sum_{t=1}^{\tmax} \sum_{ i\neq D, j } A_{ij} \mu^{t}_{ij} \nonumber \\
        && \times \bigg[ - \flux^{t}_{ij} + \density^{0}_{ij} \Poi(t-\tf_{ij}|\lambda^{0}_{ij}) \indicator_{t\ge \tf_{ij}}  \nonumber \\
        && + \sum^{t-\tf_{ij}}_{t'=1}  ( \density^{t'}_{ij}\!-\!  \density^{t'-1}_{ij} \!+\! \flux^{t'-1}_{ij} ) \Poi(t-t'-\tf_{ij}|\lambda^{t'}_{ij})  \bigg] \\
        &+& \sum_{t=1}^{\tmax} \sum_{ i\neq D, j } A_{ij} \eta^{t}_{ij} \bigg[ - \density^{t}_{ij} + p^{t-1}_{ij}\sum_{\substack{k\in\partial i \\ k\neq D}} \flux^{t-1}_{ki} + (\density^{t-1}_{ij} - \flux^{t-1}_{ij}) \bigg] \nonumber. \label{eq:Lagrangian_si}
    \end{eqnarray}}
    
    The optimal solution satisfies the optimality condition $\partial {\cal L} / \partial \{ \mu^{t}_{ij}, \eta^{t}_{ij}, \density^{t}_{ij}, \flux^{t}_{ij}, w^{t}_{ij} \} = 0$. Variation of ${\cal L}$ with respect to the Lagrange multipliers by differentiating $\partial {\cal L} / \partial \mu^{t}_{ij} = 0, \partial {\cal L} / \partial \eta^{t}_{ij} = 0$ returns the forward dynamical equations Eq.~(\ref{eq:forward1_si}) and Eq.~(\ref{eq:forward2_si}), while the variation with respect to the state variables by $\partial {\cal L} / \partial \flux^{\tau}_{mn} = 0$ and $\partial {\cal L} / \partial \density^{\tau}_{mn} = 0$ yields two sets of dual equations $\forall 1 \leq \tau < \tmax, mn\in\cE, m\neq D$
    \begin{eqnarray}
    	\mu^{\tau}_{mn} &=& \frac{1}{\sum_{e \in \cE} \density^{0}_{e} }(\tmax-\tau)\indicator_{n=D} \nonumber \\
    	&+& \sum_{t=\tau+1+\tf_{mn}}^{\tmax}\mu^{t}_{mn}  \Poi(t-\tau-1-\tf_{mn}|\lambda^{\tau+1}_{mn}) \indicator_{\tau<\tmax} \nonumber \\ 
    	&+& \sum_{j\in\partial n} \eta^{\tau+1}_{nj} p^{\tau}_{nj} \indicator_{n \neq D}\indicator_{\tau<\tmax} - \eta^{\tau+1}_{mn}\indicator_{\tau<\tmax}, \label{eq:backward1_si} 
    \end{eqnarray}
    {\small 
    \begin{eqnarray}
        \eta^{\tau}_{mn} &=& \eta^{\tau+1}_{mn} + \sum_{t=\tau+\tf_{mn}}^{\tmax} \mu^{t}_{mn} \Poi(t-\tau-\tf_{mn}|\lambda^{\tau}_{mn}) \label{eq:backward2_si}  \\ 
        &-& \sum_{t=\tau+1+\tf_{mn}}^{\tmax}\mu^{t}_{mn} \Poi(t-\tau-1-\tf_{mn}|\lambda^{\tau+1}_{mn}) \nonumber \\
        &+& \sum_{t=\tau+\tf_{mn}}^{\tmax}\mu^{t}_{mn} \big( \density^{t}_{mn} - \density^{t-1}_{mn} + \flux^{t-1}_{mn} \big) \nonumber \\
        && \qquad \times \frac{\partial \Poi(t-\tau-\tf_{mn}|\lambda^{\tau}_{mn})}{\partial \lambda^{\tau}_{mn}} \frac{\partial \lambda^{\tau}_{mn}}{\partial \density^{\tau}_{mn}} \nonumber \\
        &-& (1-n) \beta g'(\density^{\tau}_{mn}) p^{g,\tau}_{mn} \big( \eta^{\tau+1}_{mn} - \sum_{j \in \partial m} p^{g,\tau}_{mj} \eta^{\tau+1}_{mj} \big)  \sum_{\substack{k\in\partial m \\ k\neq D}}\flux^{\tau}_{km}. \nonumber
    \end{eqnarray}}
    Note that these are discrete dynamics that propagates backward in time, with the corresponding terminal condition
     {\small 
    \begin{eqnarray}
        \mu^{\tmax}_{mn} &=& 0, \label{eq:term_cond1_si} \\
        \eta^{\tmax}_{mn} &=& \mu^{\tmax}_{mn} \Poi(0|\lambda^{\tmax}_{mn}) \delta_{\tf, 0} \label{eq:term_cond2_si} \\ 
        &+& \mu^{\tmax}_{mn}(\density^{\tmax}_{mn} - \density^{\tmax-1}_{mn} + \flux^{\tmax-1}_{mn}) \frac{\partial \Poi(0|\lambda^{\tmax}_{mn})}{\partial \lambda^{\tmax}_{mn}} \frac{\partial \lambda^{\tmax}_{mn}}{\partial \density^{\tmax}_{mn}} \delta_{\tf, 0}. \nonumber
    \end{eqnarray}}
    
    Lastly, the variation with respect to the control parameter $w^{\tau}_{mn}$ by $\partial {\cal L} / \partial w^{\tau}_{mn} = 0$ gives the optimality condition,
    \begin{eqnarray}
        \frac{\partial \cL}{\partial w^{\tau}_{mn}} = - n w_{0} p^{w,\tau}_{mn} \big( \eta^{\tau+1}_{mn} - \sum_{j\in\partial m} p^{w,\tau}_{mj} \eta^{\tau+1}_{mj}  \big)  \sum_{\substack{ k\in\partial m \\ km\neq D}}\flux^{\tau}_{km} = 0.  \label{eq:dLdw_si} 
    \end{eqnarray}
    
    In principle, the above optimal control problem can be solved by iterating the equations following the procedure outlined below, until a predefined accuracy or execution time is reached
    \begin{enumerate}
        \item Start from the initial condition of Eq.~(\ref{eq:init_cond_si}), propagate the state variables forward according to Eq.~(\ref{eq:forward1_si}) and Eq.~(\ref{eq:forward2_si}).
        \item Start from the terminal condition of Eq.~(\ref{eq:term_cond1_si}) and Eq.~(\ref{eq:term_cond2_si}), propagate the Lagrange multipliers dynamics backward according to Eq.~(\ref{eq:backward1_si}) and Eq.~(\ref{eq:backward2_si}).
        \item Solve Eq.~(\ref{eq:dLdw_si}) in order to update the control variables $\{ w^{\tau}_{mn} \}$.
    \end{enumerate}
    In practice, the update in Step 3 may result in radical changes of the control variables and lead to diverging behaviors of the algorithm \cite{QianxiaoLi2017}. We find that an adoption of a gradient ascent approach to update the control variable $w^{\tau}_{mn}$ through 
    \begin{eqnarray}
        w^{\tau}_{mn} &\leftarrow& w^{\tau}_{mn} + s \frac{\partial \cL}{\partial w^{\tau}_{mn}},
    \end{eqnarray}
    is more efficient and provides a remedy to possible divergence issues. Together with Step 1 and Step 2, this update will gradually drive the system towards a locally optimal state. In practice, several different step sizes $s$ are used and different initial guesses of $\{ w^{\tau}_{mn} \}$ are given, among which the best solution is selected. We also fix the number of iterations, such that the algorithmic complexity is $O(|E|T)$, which is linear in the number of dynamical variables.

\section{Comparison to IPOPT Solver}
\begin{figure}
	\includegraphics[scale=1.2]{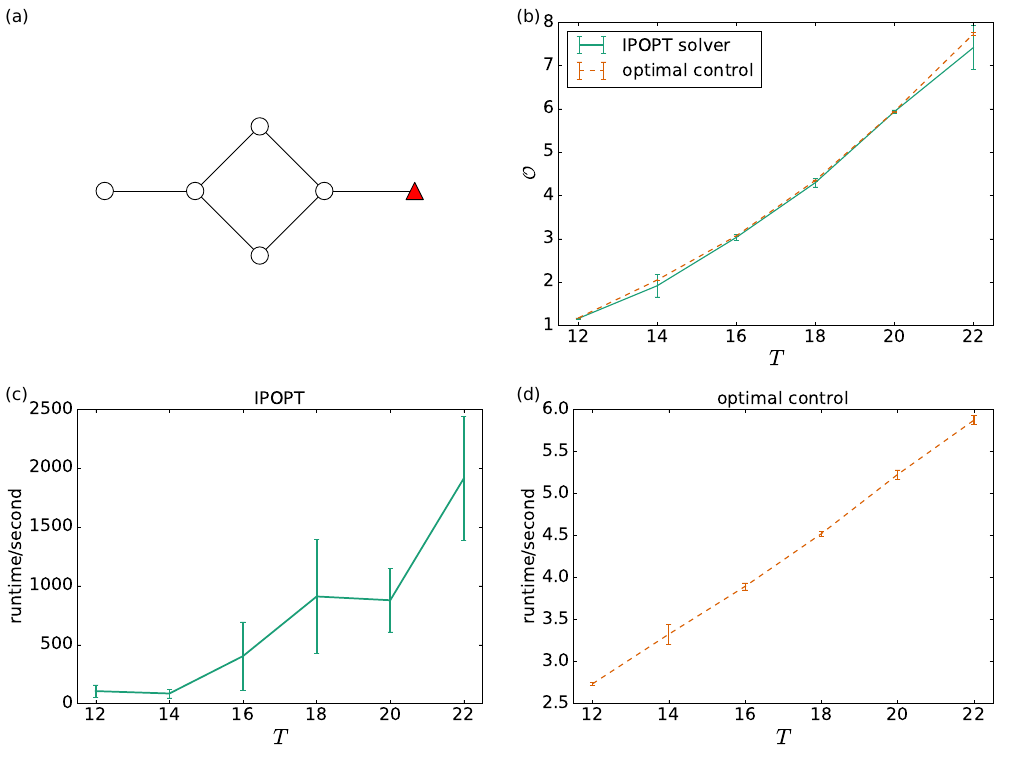}
	\caption{Comparison between the optimal control method and state-of-the-art IPOPT solver on a small network. Each data point in (b)(c)(d) is averaged over 5 converged instances starting with different random initial values of $\bm{w}$. (a) The network under consideration. The initial traffic load level is $L=\frac{1}{8}$. (b) Objective function $\mathcal{O}$ as a function of $T$. Both methods achieve similar performance. (c) Runtime of IPOPT solver as a function of $T$. (d) Runtime of optimal control method as a function of $T$.}
	\label{fig:benchmark}	
\end{figure}

In this section, we compare the optimal control method suggested in this work to the IPOPT (Interior Point Optimizer) nonlinear programming solver~\cite{Wachter2006}, implemented within the constrained optimization framework JuMP~\cite{Dunning2017}, on a small network as shown in Fig.~\ref{fig:benchmark}(a). 
The optimization problem under consideration is highly nonlinear and non-convex, with a landscape characterized by multiple local maxima. Similar to the optimal control method, different initial guesses of $\{ w^{\tau}_{mn} \}$ are provided. It is observed that the IPOPT solver does not converge with some starting values of $\{ w^{\tau}_{mn} \}$. In all the instances where IPOPT converges, the solver achieves locally optimal solution. We record the objective function values and runtime of the converged instances. The runtime is measured as the CPU runtime of the algorithms (CPU in used: Intel i5-8350U). As shown in Fig.~\ref{fig:benchmark}, the optimal control algorithm achieves similar performance as the IPOPT solver in terms of objective function values, while it is more stable and computationally efficient. 
The poor scaling property of the IPOPT solver in this problem hinders the comparison in larger networks. One possible reason for its poor scalability is that IPOPT uses second-order methods for the updates, which is much more costly in systems of large sizes per iteration than the first-order method used in the optimal control approach developed here. At the same time, a faster convergence rate of second-order methods is not guaranteed for non-convex problems considered here. 
The linear scaling of the computational cost with respect to the number of dynamical variables facilitates the application of our method to large size real-world networks.

\bibliography{reference}